\documentclass[%
reprint,
superscriptaddress,
 amsmath,amssymb,
 aps,
]{revtex4-2}

\usepackage{graphicx}
\usepackage{dcolumn}
\usepackage{bm}
\usepackage{mhchem}
\usepackage{placeins}
\usepackage{hyperref}

\newcommand{\zh}{\bm}

\newcommand{\zhg}{{\zh g}}

\newcommand{\zhq}{{\zh q}}
\newcommand{\zhr}{{\zh r}}
\newcommand{\zhu}{{\zh u}}
\newcommand{\zhv}{{\zh v}}

\newcommand{\GRECP}{{GRPP}}

\newcommand{\zhrho}{{\zh \rho}}

\begin{document}

\preprint{APS/123-QED}

\title{Multiconfigurational Analysis of Local Electronic Structure of \ce{RuO2}\\ Using Relativistic Embedded Clusters}

\author{I. A. Zhosan }
 \email{iigorezhosan@gmail.com}
\affiliation{
St. Petersburg State University, 7/9 Universitetskaya nab., St. Petersburg, 199034, Russia
}%
\affiliation{
	B. P. Konstantinov Petersburg Nuclear Physics Institute of National Research Center “Kurchatov Institute” (NRC “Kurchatov Institute”–PNPI), Gatchina, Leningrad district, 188300, Russia
}%
\affiliation{
St. Petersburg State Technological Institute (Technical University), 26 Moskovski ave., St. Petersburg, Russia, 190013
}

\author{Yu. V. Lomachuk}
\affiliation{
St. Petersburg State University, 7/9 Universitetskaya nab., St. Petersburg, 199034, Russia
}%
\affiliation{
	B. P. Konstantinov Petersburg Nuclear Physics Institute of National Research Center “Kurchatov Institute” (NRC “Kurchatov Institute”–PNPI), Gatchina, Leningrad district, 188300, Russia
}
\author{D. A. Maltsev}
\affiliation{
St. Petersburg State University, 7/9 Universitetskaya nab., St. Petersburg, 199034, Russia
}
\affiliation{
	B. P. Konstantinov Petersburg Nuclear Physics Institute of National Research Center “Kurchatov Institute” (NRC “Kurchatov Institute”–PNPI), Gatchina, Leningrad district, 188300, Russia
}
\author{I. A. Moiseev}
\affiliation{
St. Petersburg State University, 7/9 Universitetskaya nab., St. Petersburg, 199034, Russia
}
\affiliation{
	B. P. Konstantinov Petersburg Nuclear Physics Institute of National Research Center “Kurchatov Institute” (NRC “Kurchatov Institute”–PNPI), Gatchina, Leningrad district, 188300, Russia
}
\author{O. Yu. Andreev}
\affiliation{
St. Petersburg State University, 7/9 Universitetskaya nab., St. Petersburg, 199034, Russia
}
\affiliation{
	B. P. Konstantinov Petersburg Nuclear Physics Institute of National Research Center “Kurchatov Institute” (NRC “Kurchatov Institute”–PNPI), Gatchina, Leningrad district, 188300, Russia
}

\begin{abstract}
We present a multiconfigurational, relativistic embedded-cluster study of the local electronic structure of ruthenium dioxide (\ce{RuO2}), a candidate altermagnetic material. Starting from free \ce{Ru} ions, we progressively build up the local environment through a \ce{Ru+Q6} electrostatic model, a bare \ce{[RuO6]^{8-}} ligand model, and finally a high-accuracy \ce{RuO6}@CTEP embedded cluster that reproduces the crystalline surroundings. All systems are treated at the SA-CASSCF and NEVPT2+SOC levels of theory to capture strong electron correlation and spin-orbit coupling on an equal footing. While the formal local site symmetry of the \ce{Ru} sites in \ce{RuO2} is orthorhombic ($D_{2h}$), we find that the calculated $4d$-orbital energy spectrum and its splitting pattern behave much closer to the higher tetragonal ($D_{4h}$) symmetry, preserving a strong quasi-degeneracy among the relevant $4d$ orbitals. Since the local $xy$ quadrupolar order responsible for altermagnetic spin splitting in independent-particle models relies on this symmetry reduction, its suppression by orbital quasi-degeneracy offers a natural explanation for why altermagnetism is not observed in bulk \ce{RuO2} experiments, in contrast to the robust altermagnetic signatures reported in strained thin films.
\end{abstract}

\date{\today}

\maketitle


\section{\label{sec:intro}Introduction}
Altermagnetism has recently emerged as a distinct and fundamental magnetic phase of matter, disrupting long-standing conventional magnetic classifications \cite{vsmejkal2022beyond,Smejkal2022PhysRevX.12.040501,mazin2022altermagnetism}. Traditionally, magnetic materials have been viewed through a strict dichotomy: ferromagnets, which exhibit macroscopically broken time-reversal symmetry and large net magnetizations, and conventional antiferromagnets, which possess a fully compensated, zero net magnetization. In antiferromagnets, the underlying spin-up and spin-down magnetic sublattices can be perfectly mapped onto one another via a spatial translation (shift) or spatial inversion. This geometric coupling protects Kramers spin degeneracy, forcing the electronic energy bands for both spin states to remain completely degenerate throughout the entire momentum space.

While altermagnets share a zero net magnetization with antiferromagnets due to fully compensated magnetic moments, their spin-up and spin-down sublattices cannot be interconnected by pure translations or inversions. Instead, they are superimposed strictly through additional spatial rotations (such as $90^\circ$ turns), improper rotations, or specific reflection-based mirror symmetries \cite{vsmejkal2022beyond,Smejkal2022PhysRevX.12.040501,roig2024minimal,Das2026w47d-pjxf}. This unique geometric configuration lifts the Kramers spin degeneracy, inducing an anisotropic spin splitting in the electronic band structure. On a microscopic level, this unconventional time-reversal symmetry breaking allows altermagnets to host transport phenomena previously considered exclusive to ferromagnets, such as the anomalous Hall effect, while maintaining a vanishing magnetic footprint due to their zero net magnetization. Consequently, altermagnets have attracted intense interest as highly promising candidates for next-generation spintronic applications.

At the atomic scale, the mechanism driving altermagnetism originates from a subtle interplay between local and global crystal symmetries~\cite{vsmejkal2022beyond,Smejkal2022PhysRevX.12.040501,roig2024minimal,Gondolf2025PhysRevB.111.174436}. Because the local environment defined by the exact atomic site symmetry group possesses a lower symmetry than the global crystal lattice point group, local multipolar moments emerge that would otherwise be prohibited by the space lattice symmetry. Accordingly, the magnetic atoms within a single unit cell experience asymmetric, localized geometric forces. When these alternating local crystal-field distortions combine with alternating magnetic moments, the material transitions into an altermagnetic state characterized by zero net magnetization and spin-split band structures along specific crystallographic directions.

Ruthenium dioxide (\ce{RuO2}) is widely considered a highly promising candidate for hosting this unconventional altermagnetic order \cite{Jovic2018PhysRevB.98.241101,Smejkal2020Sience,Gonzalez-Hernandez2021PhysRevLett.126.127701,Feng2022Nature,vsmejkal2022beyond,Smejkal2022PhysRevX.12.040501,Hiraishi2024PhysRevLett.132.166702,Fang2024PhysRevLett.133.106701,roig2024minimal}. The site and full point-group symmetries of its crystal structure clearly predict the viability of altermagnetism. Specifically, the exact local $D_{2h}$ site symmetry of the ruthenium atoms, surrounded by a flattened and rotated octahedron of $\ce{O}$ atoms, allows for a staggered local quadrupolar order that coexists with collinear N'{e}el magnetism within the $D_{4h}$ tetragonal symmetry of the $\ce{Ru}$ sublattice. Several theoretical works have developed models to describe the altermagnetic phase in this crystal, investigating its fundamental electronic signatures and physical manifestations \cite{roig2024minimal,Gondolf2025PhysRevB.111.174436,Hiraishi2024PhysRevLett.132.166702}.

However, despite strong theoretical support, experimental investigations of \ce{RuO2} have yielded contradictory results. While some macroscopic transport measurements (such as the anomalous Hall effect) and microscopic spectroscopic techniques (such as spin-resolved photoemission spectroscopy) report evidence of altermagnetic spin splitting \cite{berlijn2017itinerant,Zhu2019PhysRevLett.122.017202,Feng2022Nature,Bose2022,Bai2022PhysRevLett.128.197202,Karube2022PhysRevLett.129.137201,He2025_NatCommun,Akashdeep2026}, other experimental works report an  absence of the effect \cite{Hiraishi2024PhysRevLett.132.166702,Kessler2024,Song2025}, leaving the true magnetic ground state of the material an open question requiring further clarification. Previous theoretical descriptions of the \ce{RuO2} crystal have been largely constrained by standard density functional theory (DFT) and DFT$+U$ methods. However, DFT$+U$ approaches suffer from inherent parameter inconsistencies, as the Hubbard $U$ term must be manually tuned to balance both the magnetic order and electronic structure of \ce{RuO2}. Furthermore, mean-field treatments fail to correctly capture the multiconfigurational multiplet structure and strong electronic correlations inherent to the open-shell \ce{Ru^{4+}} ion. To resolve these ambiguities, recent experimental efforts have focused on high-quality sample fabrication. For instance, recent studies provided compelling evidence of altermagnetism by successfully synthesizing single-variant altermagnetic \ce{RuO2}(101) thin films via fully epitaxial growth on sapphire substrates~\cite{He2025_NatCommun,Akashdeep2026}.

In this work, we present a multiscale theoretical approach to resolve these open questions.  We initially study the bulk material using periodic boundary conditions within a DFT framework. This baseline calculation is used to relax the crystal geometry, analyze the electron density, and construct the specialized non-local embedding potential required for subsequent steps. To transcend conventional DFT limitations and accurately capture local electronic correlations alongside relativistic effects at the ruthenium site, we construct finite cluster models embedded in an effective crystalline environment. Specifically, we implement the compound-tunable embedding potential (CTEP) technique \cite{maltsev2021compound,lomachuk2020compound,oleynichenko2024compound,shakhova2022compound,maltsev2025electronic,Khadeeva16022026} to build an embedded cluster model. Within this CTEP framework, we design a minimal cluster where the core \ce{RuO6} octahedron is treated as the primary quantum subsystem, while the long-range crystal environment is simulated via a combination of shape-consistent tunable pseudopotentials and optimized fractional point charges. This advanced embedding methodology allows us to bypass mean-field approximations and deploy highly accurate, multireference wavefunction-based methods, such as the Complete Active Space Self-Consistent Field (CASSCF) and the N-Electron Valence State Perturbation Theory (NEVPT2), to rigorously capture many-body correlation effects at the central ruthenium site.

Specifically, we theoretically investigate how the local deviation from $D_{4h}$ symmetry down to $D_{2h}$ symmetry qualitatively and quantitatively influences the electronic states of the crystal. This allows us to determine the fundamental parameters of the local electronic structure-crystal field splitting and spin-orbit interaction for \ce{Ru^4+} in a crystalline environment. Our \textit{ab initio} calculations provide a physical explanation for why a prominent altermagnetic order may appear absent or obscured in bulk \ce{RuO2} crystal samples. The results of this work are aimed at establishing the parameters necessary for the possible classification of the magnetic state of \ce{RuO2} as an altermagnet. Furthermore, we study the conditions required to manifest visible altermagnetism, focusing specifically on thin films, defect-engineered crystals, and systems under external fields.

\section{\label{sec:compute}Computational methodology}
The electronic structure of $\ce{RuO_{2}}$ was investigated by initially performing periodic DFT calculations of the bulk crystal. Subsequently, a cluster fragment was isolated from the optimized structure using the CTEP method. This embedded cluster model was then investigated using multiconfigurational wavefunction methods, specifically CASSCF and NEVPT2, with spin-orbit coupling (SOC) taken into account.

\subsection{\label{sec:period}Periodic calculation}
\begin{figure}[b]
  \centering
  \includegraphics[width=0.5\columnwidth]{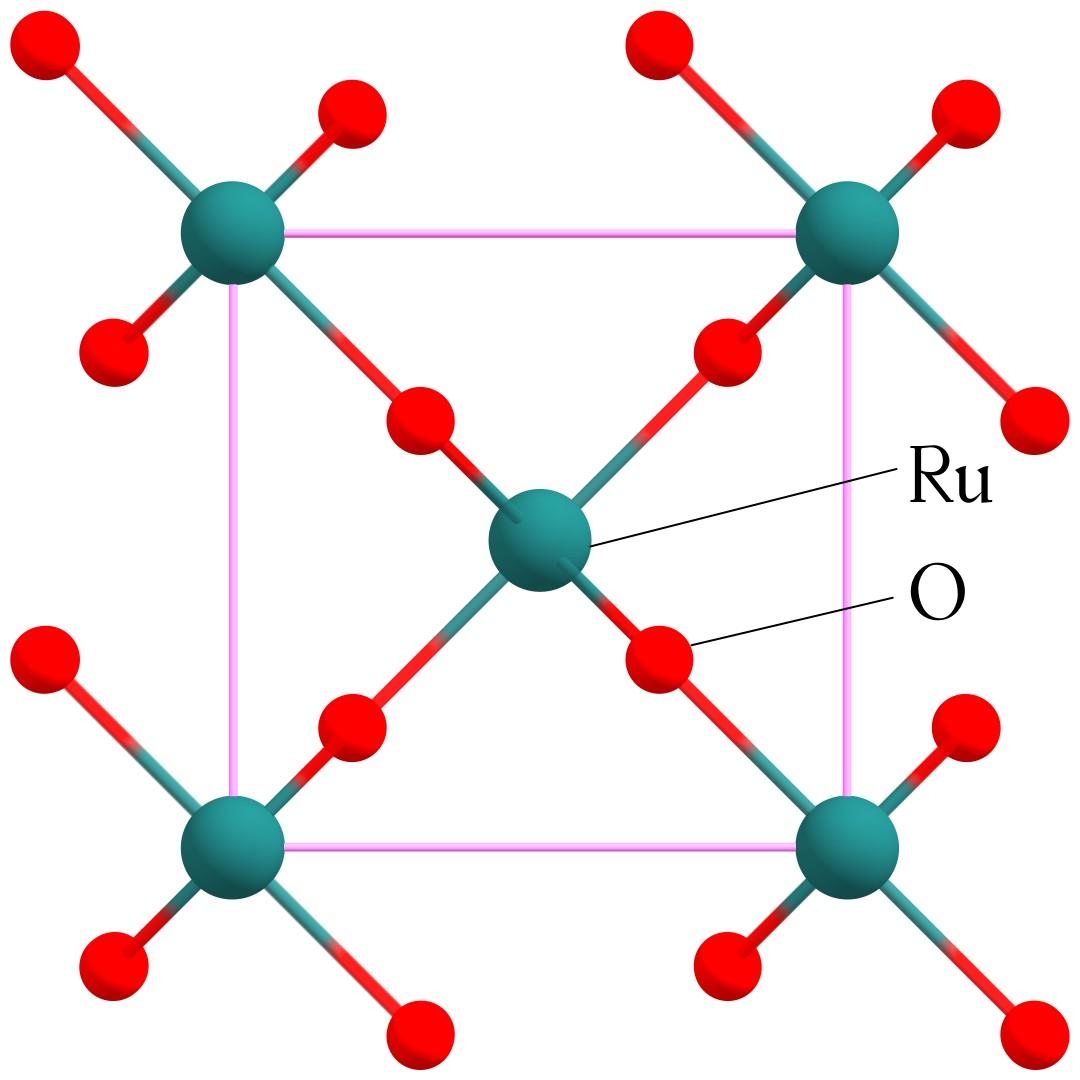}
  \caption{\label{fig:ruo2}Crystal structure of \ce{RuO2} with \ce{Ru} atoms in green and \ce{O} atoms in red. The four corner \ce{Ru} atoms form a square plane, while the central \ce{Ru} atom is displaced from this plane. Each \ce{Ru} atom is surrounded by a compressed octahedron of \ce{O} atoms (red spheres), featuring two short apical Ru--O bonds (fully shown) and four longer equatorial Ru--O bonds (only two of which are shown for clarity).}
\end{figure}

Ruthenium(IV) oxide, \ce{RuO2}, crystallizes in the rutile structure with the space group $P4_2/mnm$ (No. 136) \cite{boman1970refinement,bolzan1997structural} (structure is presented in Fig.~\ref{fig:ruo2}). To model the expected altermagnetic (AM) state, characterized by ordered antiparallel spins and a spin-split electronic structure, antiferromagnetic (AFM) ordering is specified in real space in the theoretical calculations. Simulating the AFM phase with two magnetic centers per unit cell requires a reduction in the symmetry of the crystal lattice. Therefore, all periodic computations in this work are performed within the orthorhombic space group $Cmmm$ (No. 65).

Periodic calculations were performed using the CRYSTAL software package \cite{dovesi2018crystal17,erba2022crystal23,crystal23}, which implements the CO-LCAO method, where atomic orbitals are approximated by Gaussian-type functions. To describe the ruthenium atom, the generalized relativistic pseudopotential (\GRECP) with 28-core-electrons was utilized, specifically incorporating the averaged relativistic effective potential (AREP) part to account for scalar relativistic effects and the leading part of the Breit interaction~\cite{titov1999generalized,qchem}. Special attention was paid to constructing a consistent contracted basis set of the $(7s,6p,6d,3f)/[5s,3p,3d,3f]$ type. The basis for the Ru atom was obtained by optimizing the initial def2-QZVP basis set \cite{weigend2005balanced} based on an energy-minimization criterion: the core exponents were optimized and re-contracted in the atomic calculation to match the pseudofunctions of the \GRECP. The oxygen atoms were described using an all-electron basis set of the type $(10s,6p,1d)/[4s,3p,1d]$, adopted from \cite{kuklin2022structural}.
  
The calculations were performed within the framework of DFT using a spin-polarized solution and the r2SCAN meta-GGA functional. The choice of this functional is based on recent theoretical studies of \ce{RuO2} \cite{meinert2026meta, yumnam2025constraints} and is confirmed by our own benchmark computations using available DFT functionals.

The lattice parameters, obtained through geometry optimization, are $a = b = 4.4876\,$\AA{}
and $c = 3.1065\,$\AA. These values agree well with experimental data ($a = b = 4.49\,$\AA{} and $c = 3.11\,$\AA{} \cite{bolzan1997structural}). According to DFT calculation, the ground state of the system is metallic and is characterized by spin splitting. The magnetic moment on the ruthenium atom is $0.92\,\mu_B$, where $\mu_B$ is the Bohr magneton.

\subsection{\label{sec:cluster} CTEP based cluster construction}
In this work, the CTEP method is employed to construct an embedded cluster model of the bulk \ce{RuO2} crystal, denoted as \ce{[RuO6]}@CTEP.

\begin{figure}[b]
  \includegraphics[width=0.98\columnwidth]{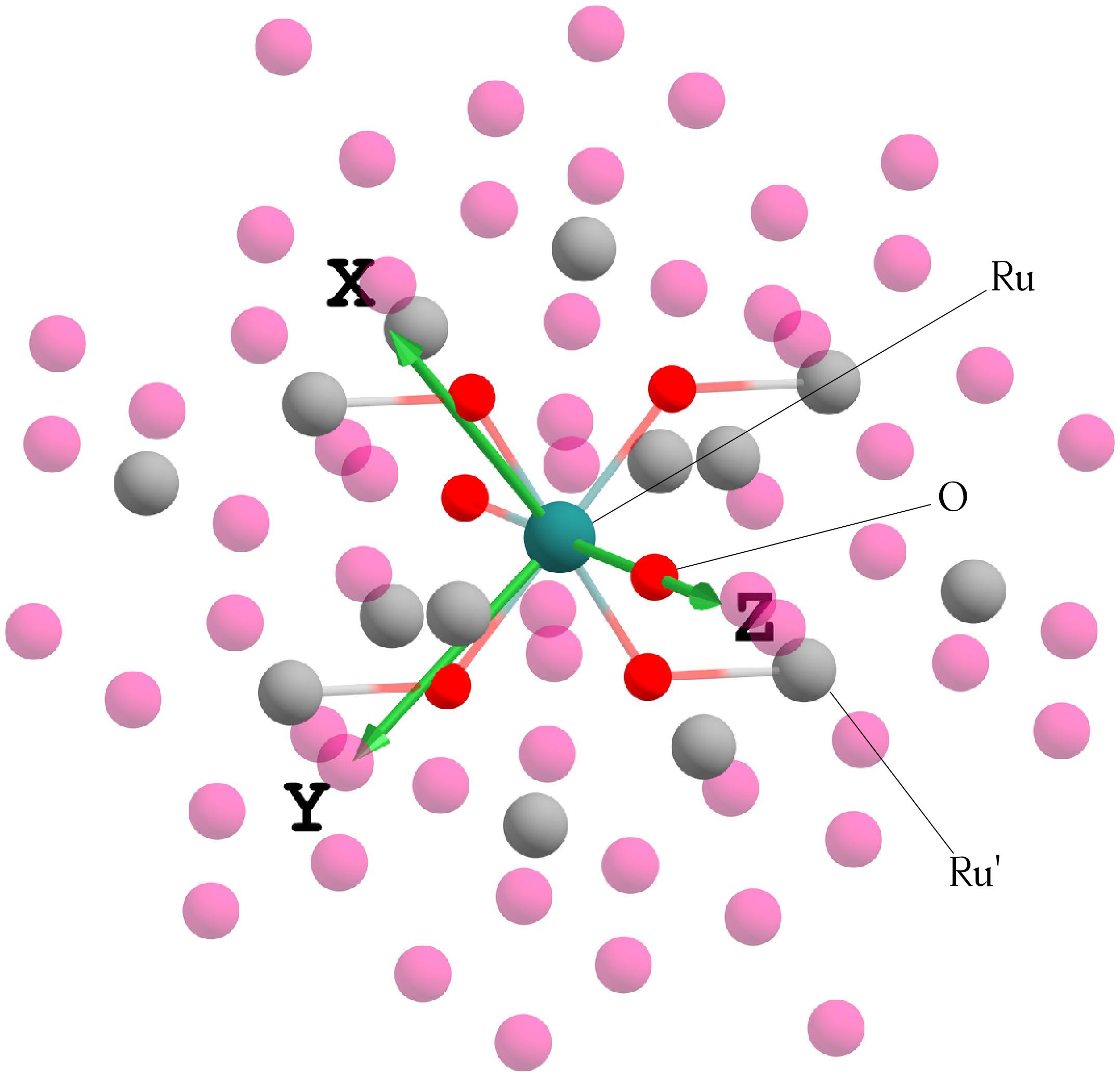}
  \caption{\label{fig:cluster} Cluster model of ruthenium dioxide (\ce{RuO2}) with the structural formula $\text{Ru}-\text{O}_6-\text{Ru}'_{14}-\text{Q}_{58}$. The pseudoatoms representing the nearest cationic environment (NCE,  $\text{Ru}'_{14}$) and the nearest anionic environment (NAE, \ce{Q_{58}}) are shown as gray and pink spheres, respectively. The central \ce{Ru} atom and the surrounding \ce{O} octahedron that form the main cluster are depicted in green and red, respectively. The octahedron exhibits two distinct Ru--O distances: four equatorial bonds to O atoms in the $xy$-plane of 1.933\,\AA, and two axial bonds along the $z$-axis of 1.988\,\AA.}
\end{figure}

The geometry of the constructed molecular cluster is identical to that optimized in the periodic DFT computation. Within the CTEP method, the cluster is divided into three distinct regions. The central region (the main cluster) is represented by a single ruthenium atom and its six nearest oxygen atoms, forming a first coordination sphere with a distorted octahedral geometry. The atoms in the main cluster are described by the same basis sets as those used in the periodic calculation. Next, the main cluster is embedded in the CTEP potential, which is constructed from the nearest cation environment (NCE) and the nearest anion environment (NAE). This potential must reproduce the behavior of the electron density obtained in the periodic calculation within the main cluster region. The choice of the NCE and NAE regions is dictated by the local symmetry of the system. In the $\ce{RuO_{2}}$ system, the ruthenium atoms occupy the 2a Wyckoff position, which corresponds to $D_{2h}$ symmetry. The complete cluster is shown in Fig.~\ref{fig:cluster}.

The NCE region includes the 14 nearest Ru atoms surrounding the main cluster, defining a radius of $3.5\,$\AA{} for the NCE region. This specific coordination sphere was selected to capture all major crystal-field effects exerted on the main cluster electrons. The atoms in the NCE region are represented as pseudoatoms (Ru$'$) with an effective charge and are modeled by a ``compound-tunable'' pseudopotential (CTPP) with a minimal $(3s,2p,2d)/[3s,2p,2d]$ basis set. The CTPP is constructed from the 28-core-electron \GRECP ~used to describe the Ru atoms in the main cluster region. Details of the CTPP construction are presented in Appendix~\ref{lc-ecp}.

The NAE region, which has a radius of $6.0$\,\AA, contains 58 oxygen atoms. This layer is defined according to heuristic criteria: the radius of the NAE region must be 1.5--2.0 times that of the NCE region. The atoms in the NAE region are modeled as fractional point charges.

The resulting \ce{[RuO6]}@CTEP cluster is electrically neutral, with the charge balance condition given by:
\begin{eqnarray}
Q_{\text{MC}}+Q_{\text{NCE}}+Q_{\text{NAE}}&=&0
\,,
\end{eqnarray}
where $Q_{\text{MC}}$, $Q_{\text{NCE}}$, and $Q_{\text{NAE}}$ are the charges of the main cluster, NCE, and NAE regions, respectively.
In the main cluster region, atoms are assigned their formal oxidation states, whereas the charges in the NCE and NAE regions are chosen to ensure the overall neutrality of the entire system.

According to the CTEP methodology, the effective charges of the NCE and NAE regions are optimized. The optimization convergence is monitored by the root-mean-square (RMS) force on the main-cluster atoms (defined the \textit{ab~initio} total-energy gradient with respect to atomic displacements):

\begin{eqnarray}
f_{\mathrm{RMS}} = \sqrt{ \frac{1}{N_{\mathrm{MC}}} \sum_{A \in \mathrm{MC}} \sum_{i=x,y,z} \left( \frac{\partial E}{\partial x_{A,i}} \right)^2 }
\,,
\end{eqnarray}
where $ N_{\mathrm{MC}}$ is the number of atoms in the main cluster. The optimization is regularized via SVD of the electrostatic response matrix. Details are provided in Appendix~\ref{svd}. Upon completion of the optimization procedure, the obtained RMS value of the forces does not exceed $1\times10^{-6}$~a.u.

\subsection{\label{sec:ma_calc}Molecular and atomic calculations}
Electronic structure calculations for the atomic and molecular systems investigated in this work, including the constructed clusters, were performed using the PySCF software package \cite{sun2020recent} integrated with the LIBGRPP library via a recently developed Python interface \cite{Oleynichenko2026librgpp,Seregin2026}. The initial molecular orbitals were generated using the restricted open-shell Hartree–Fock (ROHF) method. Relativistic and quantum electrodynamic (QED) corrections were incorporated within the \GRECP ~framework developed by the Gatchina group \cite{qchem}, employing both the averaged relativistic effective potential (AREP) and effective spin-orbit potential (ESOP) variants.

The nominal oxidation state of Ru in ruthenium dioxide is $+4$, corresponding to a $4d^4$ electronic configuration. This open-shell \(d\)-electron configuration is inherently multireference in nature, making electron correlation effects of paramount importance for an accurate description.

To account for static electron correlation, complete active space self-consistent field (CASSCF) calculations were performed. In this approach, the active-space wavefunction is expressed as a linear combination of configuration state functions (or Slater determinants) generated by distributing a specified number of active electrons among a chosen set of active orbitals. At each macro-iteration, the spin-free electronic Hamiltonian ($\hat{H}^{\text{SF}}$) is diagonalized within this active space to obtain the energies of the selected electronic states (roots) together with their configuration interaction (CI) coefficients. Because multiple states were targeted simultaneously, a State-Averaged CASSCF (SA-CASSCF) approach was employed, in which the molecular orbitals are optimized through orbital rotations to minimize the state-averaged energy, defined as the weighted sum of the individual root energies. The CI-coefficient and orbital optimizations are coupled and were iterated self-consistently until convergence in both the energy and the orbital gradient was achieved.

While SA-CASSCF yields balanced wavefunctions and zeroth-order energies for each root, it does not account for dynamic electron correlation. To recover  this contribution, NEVPT2 was applied to each spin-pure state ($S$), yielding the corrected energies $E^{\text{SF,corr}}_{S,k}$ and the associated spin-free wavefunctions $\Psi^{\text{SF}}_{S,k}$.

Then, SOC was included within the basis of many-electron states constructed from molecular orbitals that already accounted for scalar-relativistic interactions via the GRPP potential. Subsequent SOC corrections were introduced using the full configuration interaction state interaction spin-orbit (FCI-SISO) method, which was used to resolve the spin-orbit structure of the molecular levels by diagonalizing the total Hamiltonian:
%
\begin{eqnarray}
\hat{H}
&=&
\hat{H}^{\text{SF}} + \hat{P}\hat{H}^{\text{SO}}\hat{P} \,,
\end{eqnarray}
where $\hat{P}$ is the projector onto the active valence-electron space. SOC contributions from the core electrons were implicitly accounted for through the ESOP variant of the \GRECP.

The final energies and fine-structure wavefunctions were obtained by diagonalizing the effective Hamiltonian matrix \(\hat{H}^{\text{eff}}\):
\begin{eqnarray}
\hat{H}^{\text{eff}}_{ij}
&=&
E_{S_i,i}^{\text{SF,corr}}\delta_{S_i,S_j}\delta_{ij} + \langle \Psi^{\text{SF}}_{S_i,i} |\hat{P} \hat{H}^{\text{SO}}\hat{P} | \Psi^{\text{SF}}_{S_j,j} \rangle
\,.
\end{eqnarray}

\subsection{\label{sec:ru-2-4} Electronic structure of Ru ions}
To validate the employed approaches, we calculated the electronic structures of the \ce{Ru^2+} ($4d^{6}$ configuration) and \ce{Ru^4+} ($4d^{4}$ configuration) atomic-like systems. For the first ion, experimental spin-orbit splitting energies are available for the fine-structure levels. This allows us to verify the accuracy of our SOC calculations performed with the \GRECP ~and the chosen basis set. Since both ions have the ${}^{5}D$ term in their ground state, a direct comparison of the results is possible. Calculations using the \GRECP ~were also compared with those employing all-electron basis sets from the ANO-RCC (atomic natural orbitals relativistic core-correlated) series. The active space included only the $4d$ orbitals of ruthenium, with SA-CASSCF(6e,5o) for \ce{Ru^2+} and SA-CASSCF(4e,5o) for \ce{Ru^4+}. The number of roots in the SA-CASSCF procedure was set to cover the complete set of terms arising from the given configurations.

\subsubsection{The \ce{Ru^{2+}} ion}
The energy levels of the \ce{Ru^{2+}} ion, including its fine structure, are well documented in the literature (e.g., the NIST atomic spectra database). The SOC calculations for \ce{Ru^{2+}} were performed using the FCI-SISO method. To quantify the errors introduced by the approximate treatment of SOC in this approach, corresponding benchmark calculations were carried out.

\begin{figure}[b]
  \centering
  \includegraphics[width=0.95\columnwidth]{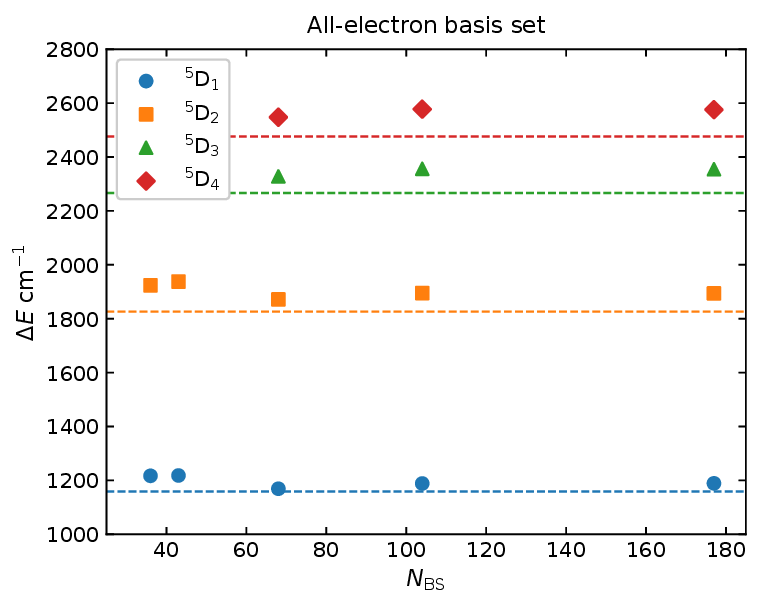}
  \vspace{0.2cm}
  \includegraphics[width=0.95\columnwidth]{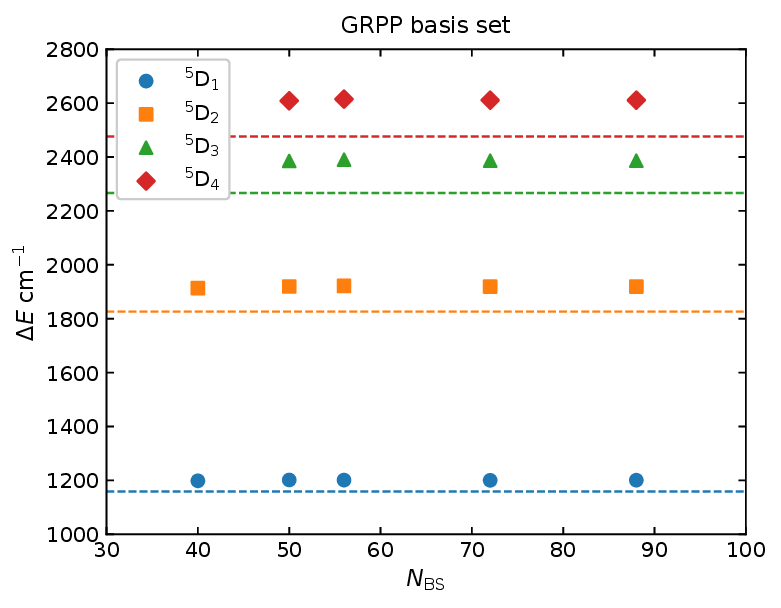}
  \caption{\label{fig:ru2}Convergence plots for the energies of the $^5D_J$ terms of the $4d^6$ configuration as a function of the basis-set size ($N_{\text{BS}}$). The energy levels appear in descending order: $J=3,2,1,0$ (representing an inverted multiplet). The ground-state $^5D_4$ level is taken as the reference baseline and is omitted from the curves. Experimental values are indicated by dashed lines.}
\end{figure}

Figure~\ref{fig:ru2} shows the basis-set energy convergence plots for \ce{Ru^{2+}}.
The convergence curves were extrapolated using a power-law expression:
\begin{eqnarray}
\label{bas}
E(N_{\text{BS}}) &=& E_{\infty} + A N_{\text{BS}}^{-\alpha},
\end{eqnarray}
where $A$ and $\alpha$ are fitting parameters determined separately for each level,
and $N_{\text{BS}}$ is the number of basis functions, which ranges from 36
to 177 for the ANO-RCC series and from 40 to 88 for the \GRECP ~basis. The
resulting extrapolation limits are illustrated alongside the raw data points
in Fig.~\ref{fig:ru2}.

The final values for the energy of the $^5 D_J$ terms (represented by $E_{\infty}$) are summarized in Table~\ref{tab:table1}. The ANO-RCC series exhibits smooth convergence and good agreement with experimental data. We therefore expect that this basis set is also suitable for the subsequent calculations on the~\ce{Ru^{4+}} ion.

Table~\ref{tab:table1} also lists the values of the effective SOC constant $\lambda$ and the corresponding one-electron parameter $\zeta_{nl}$ derived using the Landé interval rule:
\begin{eqnarray}
E(J)-E(J-1)
&=&\label{socpar}
\lambda \cdot J
\,,\quad
\lambda =\pm \frac{\zeta _{nl}}{2S}
\,.
\end{eqnarray}

\begin{table*}
\caption{\label{tab:table1} Term energies of the $^5D_J$ fine-structure levels of the $4d^6$ configuration, referenced to the lowest-energy level, in cm$^{-1}$. Results are given for all-electron calculations (``All-electron''), the \GRECP ~relativistic pseudopotential approach (``\GRECP''), and experiment (``Exp-t''). The $\Delta$ columns list the absolute deviations of the calculated values from experiment. The parameters $\lambda$ and $\zeta_{nl}$, together with their weighted averages $\lambda_{\text{avg}}$ and $\zeta_{\text{avg}}$, are the effective SOC constants and the corresponding one-electron parameters, respectively, defined via Eq.~\eqref{bas}.}
\begin{ruledtabular}
\begin{tabular}{l|cc|ccc|ccc}
Term                   & Exp-t  & $\lambda$ & All-electron & $\Delta$ & $\lambda$ & \GRECP  & $\Delta$ & $\lambda$ \\  \colrule
$^5D_4$                & 0.0    & -         & 0.0     & 0.0      & -         & 0.0    & 0.0      & -         \\
$^5D_3$                & 1158.8 & -289.7    & 1182.7  & 23.9     & -295.7    & 1200.5 & 41.7     & -300.1    \\
$^5D_2$                & 1826.3 & -222.5    & 1886.6  & 60.3     & -234.6    & 1918.6 & 92.3     & -239.4    \\
$^5D_1$                & 2266.3 & -220.0    & 2346.0  & 79.7     & -229.7    & 2384.9 & 118.6   & -233.1    \\
$^5D_0$                & 2476.0 & -209.7    & 2566.6  & 90.6     & -220.6    & 2608.9 & 132.9   & -224.0    \\  \colrule
$\lambda_{\text{avg}}$ & -      & -250.6    & -       & -        & -259.5    & -      & -        & -263.8    \\
$\zeta_{\text{avg}}$   & -      & 1002.5    & -       & -        & 1038.1    & -      & -        & 1055.3  
\end{tabular}
\end{ruledtabular}
\end{table*}

The \GRECP ~basis sets yield a similar convergence trend, although their absolute agreement with experiment is slightly poorer, as the computed state energies consistently overestimate the experimental values. This discrepancy is expected, since the basis set was optimized specifically for the higher valence states of the \ce{Ru^{4+}} ion. Nevertheless, the total spin-orbit splitting is well reproduced by both approaches, confirming the validity of the SOC treatment within the framework of the chosen method.

\subsubsection{The \ce{Ru^{4+}} ion}
To the best of our knowledge, no experimental data are available for the energy spectrum of the $\ce{Ru^{4+}}$ ion. Therefore, we utilized extrapolated values obtained with the ANO-RCC all-electron basis set to validate our remaining computational approaches. The convergence curves were extrapolated using the power law defined in Eq.~\eqref{bas}.

Table~\ref{tab:table2} presents the calculated ${}^{5}D_{J}$ multiplet terms, represented by the extrapolated limit $E_{\infty }$ (see Eq.~\eqref{bas}), along with the effective SOC constant $\lambda$ and the corresponding one-electron SOC parameter $\zeta_{nl}$ (see Eq.~\eqref{socpar}). The all-electron ANO-RCC and the \GRECP ~relativistic pseudopotential approaches yield highly consistent results. The energy deviations do not exceed $0.05\,\text{eV}$, which represents good agreement for $d$-block transition metals.

The averaged effective SOC parameters ($\lambda_{\text{avg}}$ and $\zeta_{\text{avg}}$) are consistent across both approaches and agree well with known theoretical estimates~\cite{occhialini2021local}. We observe a steady decrease in $\lambda$ when moving from the lower to the upper multiplet intervals, which reflects the clear effect of intermediate coupling and indicates a departure from pure $LS$ coupling scheme. The calculations correctly reproduce this decrease in the SOC parameters. These results confirm the adequacy of the proposed multiconfigurational approach in describing both SOC and electron correlation effects.

\begin{table*}
\caption{\label{tab:table2} Term energies of the $^5D_J$ fine-structure levels of the $4d^4$ configuration, referenced to the lowest-energy level, in cm$^{-1}$. Results are given for all-electron calculations (``All-electron'') and the \GRECP ~relativistic pseudopotential approach (``\GRECP''). The $\Delta$ column lists the absolute deviations of the \GRECP ~values from the all-electron baseline. The parameters $\lambda$ and $\zeta_{nl}$, together with their weighted averages $\lambda_{\text{avg}}$ and $\zeta_{\text{avg}}$, are the effective SOC constants and the corresponding one-electron parameters, respectively, defined via Eq.~\eqref{bas}. The parameters $A$ and $\alpha$ are power-law fitting parameters, with $\alpha$ dimensionless and all other quantities given in cm$^{-1}$.}
\begin{ruledtabular}
\begin{tabular}{l|cccc|ccccc}
Term                   & All-electron & A                   & $\alpha$ & $\lambda$ & \GRECP  & A                    & $\alpha$ & $\lambda$ & $\Delta$ \\ \colrule
$^5D_0$                & 0.0     & -                   & -        & -         & 0.0    &                      &          & -         & 0.0      \\
$^5D_1$                & 486.1   & $3.330\times10^{7}$ & 3.7813   & 486.1     & 488.4  & $3.612\times10^{4}$  & 9.9856   & 488.4     & 2.3      \\
$^5D_2$                & 1298.7  & $2.056\times10^{7}$ & 3.3931   & 406.3     & 1314.2 & $4.320\times10^{4}$  & 6.2517   & 412.9     & 15.4     \\
$^5D_3$                & 2311.8  & $1.041\times10^{7}$ & 3.0578   & 337.7     & 2352.4 & $1.000\times10^{12}$ & 7.0852   & 346.1     & 40.7     \\
$^5D_4$                & 3454.9  & $5.425\times10^{6}$ & 2.7758   & 285.8     & 3532.1 & $1.000\times10^{21}$ & 6.8043   & 294.9     & 77.2     \\ \colrule
$\lambda_{\text{avg}}$ & -       & -                   & -        & 351.1     & -      & -                    & -        & 358.6     & -        \\
$\zeta_{\text{avg}}$   & -       & -                   & -        & 1404.3    & -      & -                    & -        & 1434.4    & -      
\end{tabular}
\end{ruledtabular}
\end{table*}

We further verified that the energies obtained with the \GRECP ~pseudopotential remain almost insensitive to the addition of diffuse functions. Consequently, for the subsequent molecular cluster calculations, we employed a basis set identical to that used in the periodic calculations.

\begin{figure}[b]
  \centering
  \includegraphics[width=0.95\columnwidth]{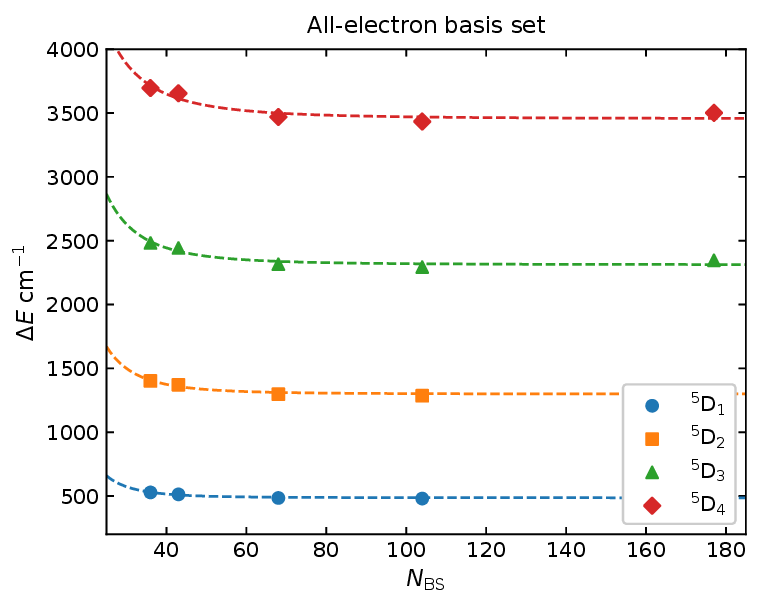}
  \vspace{0.2cm}
  \includegraphics[width=0.95\columnwidth]{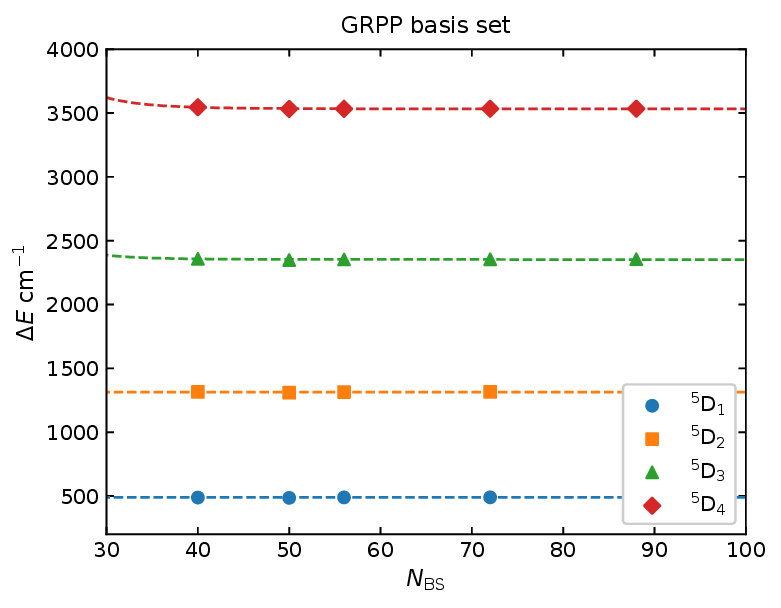}
\caption{\label{fig:ru4} Convergence plots for the energies of the $^5D_J$ terms of the $4d^4$ configuration as a function of the basis-set size ($N_{\text{BS}}$). The energy levels appear in ascending order: $J=1,2,3,4$ (representing an normal multiplet). The ground-state $^5D_0$ level is taken as the reference baseline and is omitted from the curves. The power-law fit of the all-electron basis set energies is shown as a dashed line.}
\end{figure}

\section{Results}
We investigate the energy levels of various model compounds of \ce{Ru} in a crystalline environment. Particular attention is paid to the independent-particle approximation, which is widely used for modeling the magnetic structure.

We start with the analysis of the \ce{Ru^{4+}} ion embedded in an octahedron formed by six point charges. At this stage (Sec.~\ref{sec:ru6q}), we study the influence of the point-charge values on the energy levels of \ce{Ru^{4+}}, as well as the changes in the energy structure as the symmetry of the system is reduced from $O_h$ to $D_{4h}$ and then to $D_{2h}$.

In the next stage (Sec.~\ref{sec:ruo6}), we analyze the energy structure of the $[\ce{RuO6}]^{8-}$ complex. Specifically, we investigate how the energy levels shift upon distorting the $\ce{O6}$ octahedron, which corresponds to a symmetry descent from $\mathrm{O}_h$ to $\mathrm{D}_{4h}$ and further down to $\mathrm{D}_{2h}$.

Finally (Sec.~\ref{sec:ruo6_ctep}), we examine the energy structure of the \ce{RuO_6} cluster constructed using the CTEP technique, which accurately represents a fragment of the bulk \ce{RuO_2} crystal.

\subsection{\label{sec:ru6q} The \ce{Ru^{4+} + 6Q} electrostatic model}
A simple way to emulate the crystalline environment of the \ce{Ru^{4+}} cation in \ce{RuO2} is to surround it with six equivalent point charges $Q$, forming an octahedron centered on the Ru atom. The spatial arrangement replicates the nearest-neighbor oxygen coordination shell of the \ce{RuO2} lattice. By systematically tuning both the geometry and the magnitude of $Q$, we simulate the response of the $4d^4$ configuration to crystal fields of varying strength and anisotropy. The geometry of the six point charges is varied from a regular octahedron ($O_h$), through a tetragonally compressed octahedron along the $z$-axis ($D_{4h}$), to a orthorhombically distorted arrangement ($D_{2h}$). The latter is obtained by transforming the central square (in the $xy$-plane) into a rectangle, consistent with the experimental geometry of the \ce{RuO2} crystal. The computed many-electron term energies are compiled into Tanabe--Sugano-type diagrams.

We begin by examining the electrostatic splitting of the $4d$ orbitals of \ce{Ru} in this crystalline environment. The one-electron orbital energies are obtained from a decomposition of the molecular orbitals over atomic orbitals within a CAS(4e,5o) calculation. Computational details are given in Sec.~\ref{sec:ma_calc}.

\begin{figure*}[t]
  \centering
  \includegraphics[width=0.32\textwidth]{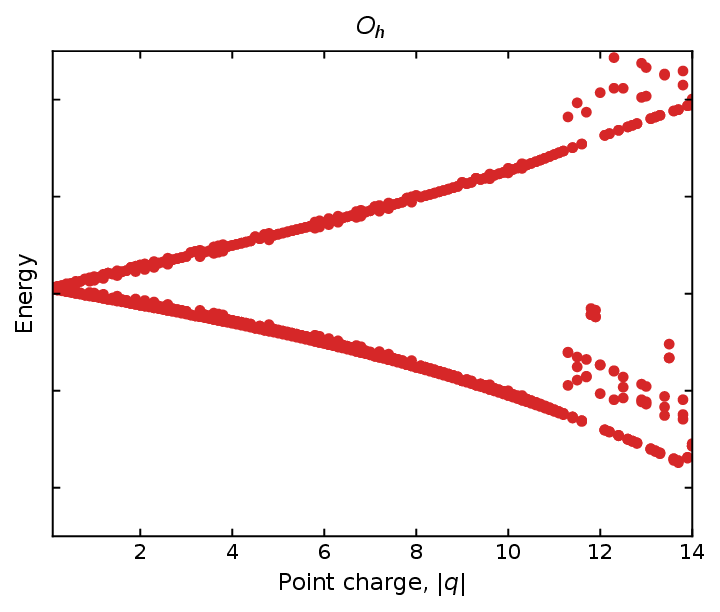}
  \hfill
  \includegraphics[width=0.32\textwidth]{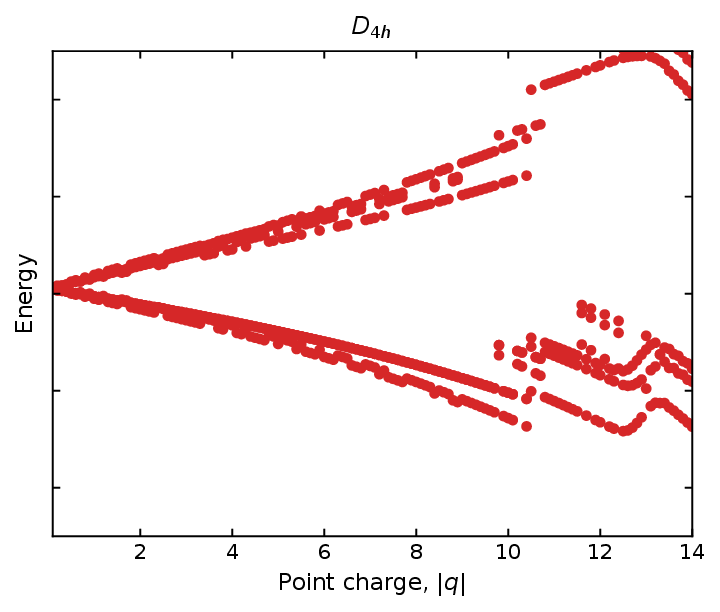}
  \hfill
  \includegraphics[width=0.32\textwidth]{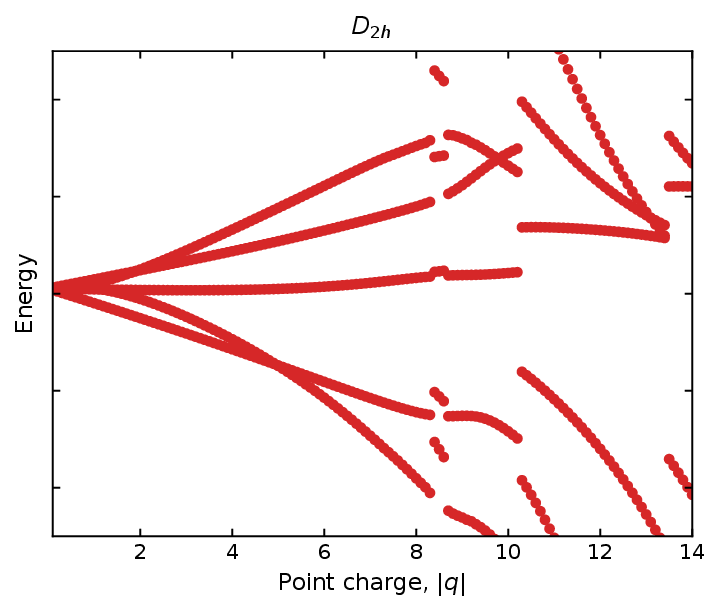}
\caption{\label{fig:mo_cas} Orbital energies of the \ce{Ru^{4+} + 6Q} system derived from the atomic Ru $4d$ orbitals, plotted as a function of the point charge magnitude $|q|$. The orbital energies are calculated at the SA-CAS(4e,5o) level and are given in arbitrary units.}
\end{figure*}

Figure~\ref{fig:mo_cas} presents these $d$-orbital energies of \ce{Ru^{4+}} as a function of the absolute point-charge value $|q|$ for each symmetry representation.

The upper panel illustrates the $O_h$ case. At $|q|=0$, all five $d$-orbitals remain degenerate. Increasing $|q|$ lifts this degeneracy, yielding the familiar $e_g$ and $t_{2g}$ manifolds. The diagram, however, also exposes an inherent limitation of the point-charge model: beyond $|q| \approx 11$, the $4d$ states are artificially destabilized by excessive electrostatic repulsion, and the splitting pattern loses its physical meaning. Since the energy splitting is proportional to $|q|$, this linear relationship will be used below when presenting the Tanabe--Sugano diagrams.

The middle panel corresponds to the $D_{4h}$ geometry (flattened octahedron). Here the degeneracy is resolved into four distinct levels: three non-degenerate states and one $e_g$ doublet, consistent with the group-theoretical decomposition of the $d$ manifold under $D_{4h}$. The model remains applicable up to $|q| \approx 10$.

The lower panel displays the $D_{2h}$ results. In this lower-symmetry environment, the orbital degeneracy is fully lifted, yielding five distinct energy levels. Notably, $D_{2h}$ symmetry allows crossings between states of different irreducible representations, giving rise to a non-trivial level structure. The applicable range contracts further, with the model breaking down beyond $|q| \approx 8$. Moreover, the overall splitting width in the $D_{2h}$ case substantially exceeds that of the $D_{4h}$ geometry, reflecting the enhanced ligand-field anisotropy.

\begin{figure*}[t]
  \centering
  \includegraphics[width=0.32\textwidth]{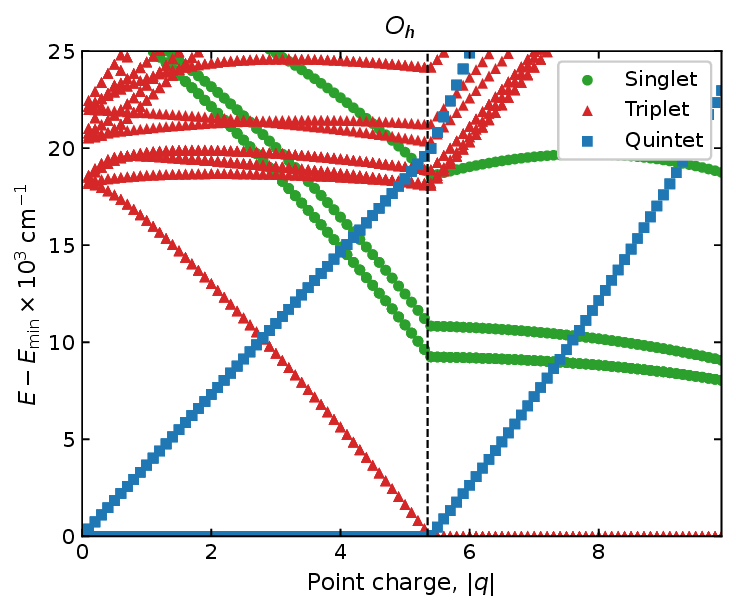}
  \hfill
  \includegraphics[width=0.32\textwidth]{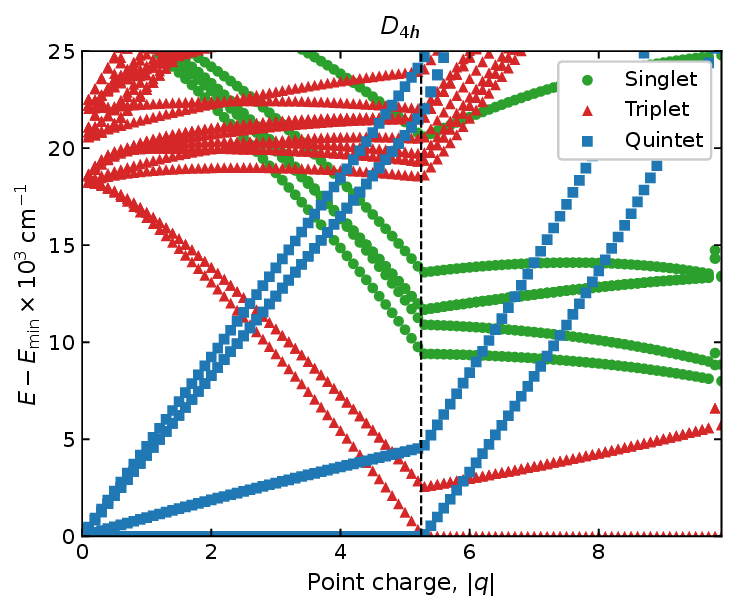}
  \hfill
  \includegraphics[width=0.32\textwidth]{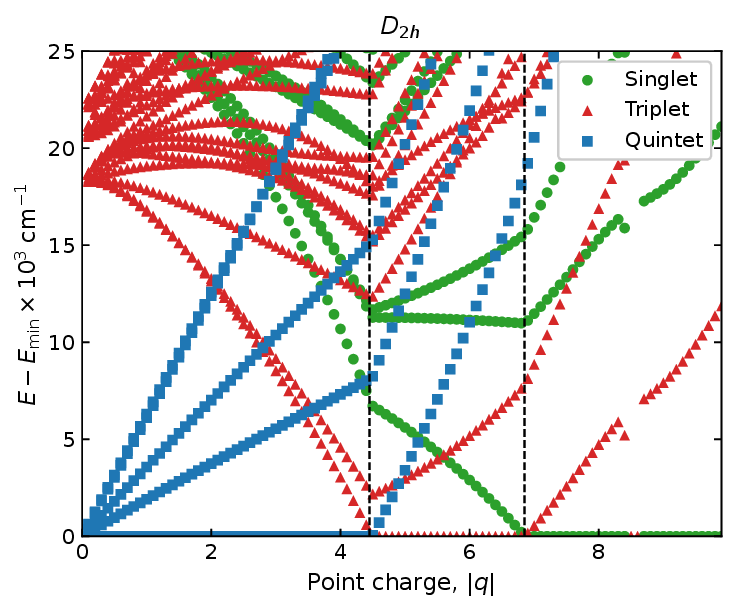}
\caption{\label{fig:nevpt2_en} Term energies of the \ce{Ru^{4+} + 6Q} system plotted as a function of the point charge magnitude $|q|$, calculated at the SA-CASSCF+NEVPT2 level of theory. Vertical dashed lines indicate the values of $|q|$ where the ground state changes. Energies are reported in \text{cm}$^{-1}$ and referenced to the lowest-energy state ($E_{\text{min}}$).}
\end{figure*}

We next investigate the many-electron term energies of the \ce{Ru^{4+} + 6Q} system, which represents the electrostatic model for the \ce{RuO2} crystal. Analysis of the calculated total-energy spectrum within this framework allows us to track changes in the excitation spectrum of the local electronic structure as the symmetry is successively reduced. Figure~\ref{fig:nevpt2_en} presents these term energies as functions of $|q|$.

The upper panel of Fig.~\ref{fig:nevpt2_en} shows the energy terms for $O_h$ symmetry. Tanabe--Sugano diagrams are conventionally plotted as a function of the one-electron $d$-orbital splitting $\Delta_0$. In the highly symmetric $O_h$ field, $\Delta_0 \propto |q|$ (see Fig.~\ref{fig:mo_cas}); therefore, the diagram presented here is effectively a genuine Tanabe--Sugano diagram, and agrees well with previous studies~\cite{tanabe1954absorption}.

For the $D_{4h}$ and $D_{2h}$ symmetries, the $d$-orbitals exhibit distinct splittings that cannot be described by a single parameter. We therefore present the energy terms directly as functions of $|q|$.

For a weak external field (small $|q|$), the ground state is a quintet. At $|q|=0$, this state is five-fold degenerate; it splits into 2, 3, and 5 components under the $O_h$, $D_{4h}$, and $D_{2h}$ fields, respectively. As $|q|$ increases, the ordering of the energy terms changes substantially: the ground state becomes a triplet for $O_h$ and $D_{4h}$ symmetry, and a singlet for $D_{2h}$ symmetry. The values of $|q|$ at which these ground-state changes occur are indicated by vertical black lines.

For the tetragonal distortion from $O_h$ to $D_{4h}$, we observe a significant increase in the energy gaps between levels. States whose splitting in the $O_h$ field was determined exclusively by spin--orbit interactions become well separated in energy due to the partial lifting of orbital degeneracy. In the strong-field regime, the first excited triplet states lie considerably higher in energy than in the cubic $O_h$ case. Nevertheless, the overall topology of the Tanabe--Sugano diagrams for the $O_h$ and $D_{4h}$ fields remains comparable.

The picture changes significantly upon further symmetry reduction to the orthorhombic $D_{2h}$ group. In the strong-field regime, the ground state is no longer a triplet but a singlet, and the range over which the triplet state remains the ground state narrows considerably. The first excited levels lie very close in energy to states of different multiplicity (singlets and quintets). This result requires cautious interpretation in view of the limitations of the electrostatic model: the point-charge approach does not account for contributions from oxygen orbitals or for polarization from neighboring ruthenium atoms in the \ce{RuO2} lattice.

The resulting diagrams can be used to map the term energies onto the symmetry and strength of the effective external field experienced by Ru in the crystal, as needed for modeling the magnetic structure.

\subsection{\label{sec:ruo6} The \ce{[RuO6]^{8-}} ligand model}
The next model, which we consider here, provides a more realistic representation of the $\ce{RuO2}$ crystal environment. It consists of a $\ce{[RuO6]^{8-}}$ complex, in which the octahedron is formed by six $\ce{O}$ atoms surrounding a central $\ce{Ru}$ atom. This extends the previous point-charge model (Sec.~\ref{sec:ru6q}) by going beyond the electrostatic approximation. The total charge of the $\ce{[RuO6]^{8-}}$ complex, determined by standard oxidation states, is $-8$.

The main objective of this model is to investigate the changes in orbital energies and the ground state upon reducing the symmetry from $\mathrm{D}_{4h}$ to $\mathrm{D}_{2h}$. As in the previous model (Sec.~\ref{sec:ru6q}), the transition from $\mathrm{O}_h$ to $\mathrm{D}_{4h}$ symmetry is achieved through tetragonal compression along the $z$-axis. The four $\ce{O}$ atoms forming the square in the $xy$-plane are arranged such that the $x$ and $y$ axes pass through opposite $\ce{O}$ atoms. The reduction from $\mathrm{D}_{4h}$ to $\mathrm{D}_{2h}$ is then realized by distorting this square into a rectangle, which incorporates the experimental geometry of the $\ce{RuO2}$ crystal. This rectangle can be characterized by the angle $\phi$ between its diagonals; $\mathrm{D}_{4h}$ symmetry corresponds to $\phi = 90^\circ$, whereas the experimental geometry corresponds to $\phi = 102.7^\circ$.

\begin{figure}[b]
  \centering
  \includegraphics[width=0.98\columnwidth]{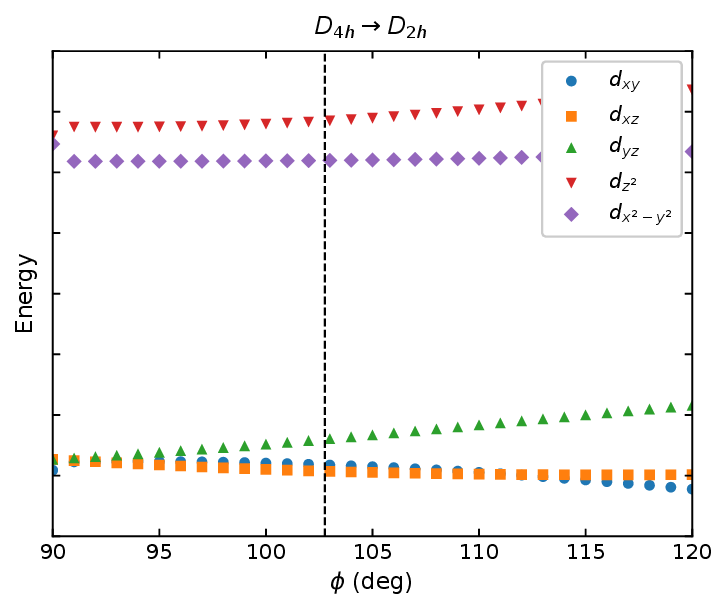}
\caption{\label{fig:cas_mo_symm} Orbital energies of the \ce{[RuO6]^{8-}} system derived from the atomic Ru $4d$ orbitals, plotted as a function of the angle $\phi$ between the diagonals of the distorted octahedron in the $xy$ equatorial plane, showing the transition from $D_{4h}$ (at $\phi=90^\circ$) to $D_{2h}$ symmetry. The vertical dashed line indicates the angle corresponding to the experimental geometry. The energies are given in arbitrary units.}
\end{figure}

The gradual deviation of $\phi$ from $90^\circ$ manifests itself in a clearly trackable splitting between the $d_{xz}$ and $d_{yz}$ orbitals, obtained via decomposition of molecular orbitals into an atomic orbital basis. The energy splitting of the $d$-electron orbitals as a function of $\phi$ is presented in Fig.~\ref{fig:cas_mo_symm}. At $\phi = 90^\circ$ ($\mathrm{D}_{4h}$ symmetry), $d_{xz}$, and $d_{yz}$ orbitals are degenerate while the $d_{z^2}$ and $d_{x^2-y^2}$ orbitals are not degenerate. The present $d_{xz}$, and $d_{yz}$ degeneracy is lifted upon lowering the symmetry.

The one-electron energies and corresponding molecular orbitals were obtained at the SA-CASSCF level of theory. To map the obtained molecular orbitals onto the d‑orbitals of the \ce{Ru^{4+}} atomic-like system, we examined the expansion coefficients of the molecular orbitals in the atomic orbital basis set. The molecular orbitals corresponding to the atomic d‑orbitals were unambiguously identified based on the dominant weight of the respective atomic $d$‑orbital contribution. The energies of these molecular orbitals are referred to as the modified atomic $d$‑orbital energies. The dependence of these energies on the octahedral rotation angle is presented in Fig.~\ref{fig:cas_mo_symm}.

To derive the crystal field splitting parameters and analyze the influence of dynamic correlation and SOC for the \ce{[RuO6]^{8-}} model at the SA-CASSCF+NEVPT2+SOC level of theory, the quintet of many-electron states ($S=2$) is considered in detail within the framework of an independent-particle interpretation. Since the wavefunction of the quintet states obtained at the SA-CASSCF level is dominated by a single determinant, this justifies the use of the independent-particle  approach for qualitative interpretation. Indeed, for the $4d^4$ configuration, five energy levels can be unambiguously identified within the quintet multiplet, each of which physically corresponds to the localization of an electron vacancy (hole) on a specific $d$-orbital. The application of group theory enables the identification of many-electron energy terms under decreasing symmetry, which is discussed in more detail in Appendix~\ref{sec:sym}. However, it should be reiterated that the parameters obtained in this way provide only a qualitative estimate, since they are derived within the framework of an independent-particle interpretation, which neglects the multiconfigurational nature of the many-electron states under study.

The details of the transformation of the term energies are presented in Table~\ref{tab:table3}. It shows the energy splitting values for the initial tetragonal geometry ($D_{4h}$) and the distorted orthorhombic structure ($D_{2h}$), whose parameters are close to the experimental bond angles in the $\ce{RuO_{2}}$ lattice ($103^{\circ } $ and $77^{\circ }$). The data presented were obtained from a many-electron spectrum calculated using the SA-CASSCF+NEVPT2+SOC level of theory. For comparison with classical models, the centers of mass of the spin-orbital multiplets are given. Explicit inclusion of SOC shows that for individual levels, the splitting is on the order of 300~$\text{ cm}^{-1}$, which is comparable to the SOC constant of the free \ce{Ru^{4+}} ion (see Section~\ref{sec:ru-2-4}).

In a spherically symmetric field, the five $d$-orbitals ($d_{z^2}$, $d_{x^2-y^2}$, $d_{xy}$, $d_{xz}$, and $d_{yz}$) are degenerate and form a set which realizes an irreducible representation of the $SO(3)$ group. Reduction of the symmetry of the external field to $O_{h}$ splits the set of five $d$-orbitals into two sets, 
$e_g$ ($d_{z^2}$, $d_{x^2-y^2}$) and $t_{2g}$ ($d_{xy}$, $d_{xz}$, and $d_{yz}$), which realize irreducible representations of the $O_h$ group. Further reduction 
of the symmetry to $D_{4h}$ splits the $e_g$ set into the $a_{1g}$ ($d_{z^2}$) and $b_{1g}$ ($d_{x^2-y^2}$) one-dimensional sets, and the $t_{2g}$ set into the 
$b_{2g}$ ($d_{xy}$) one-dimensional set and the $e_g$ ($d_{xz}, d_{yz}$) two-dimensional set. A further reduction of the symmetry to the $D_{2h}$ 
group completely lifts all remaining degeneracies, splitting the $D_{4h}$ $e_g$ set ($d_{xz}, d_{yz}$) into two one-dimensional sets, $b_{2g}$ and $b_{3g}$, 
while the $D_{4h}$ $b_{2g}$ set ($d_{xy}$) transforms into a $b_{1g}$ one-dimensional set. Finally, both the $D_{4h}$ $a_{1g}$ ($d_{z^2}$) and $b_{1g}$  ($d_{x^2-y^2}$) sets map onto the totally symmetric $a_g$ representation in $D_{2h}$.

The symmetry reduction tables in group theory show how the four-electron quintet states are composed of one-electron states from each irreducible set. The quintet states are shown in the first column of Table~\ref{tab:table3}, the components of the states (configurations) are given in the second column, and the population of each one-electron state is indicated by its exponent.

\begin{table}[b]
\caption{\label{tab:table3} Quintet terms for the \ce{[RuO_{6}]^{8-}} system in fields of $D_{4h}$ ($\phi=90^\circ$) and $D_{2h}$ (at the experimental angle $\phi$) symmetry. The first column gives the term symbol. The second column shows the corresponding electron configuration derived from the atomic Ru $4d^4$ configuration and assigned within the independent-particle approximation (the superscripts indicate the orbital population). The third column provides the energies in~$\text{cm}^{-1}$ relative to the lowest quintet state of the respective point group.}
	\begin{ruledtabular}
	\begin{tabular}{ccc}
		\multicolumn{3}{c}{$D_{4h}$}                                                                                                                        \\  \colrule
		Energy term    & Configuration                                 & Energy  \\
		${}^{5}A_{1g}$ & $(b_{2g})^1(e_{g})^2(b_{1g})^1(a_{1g})^0$          & 0.0                                       \\
		${}^{5}B_{1g}$ & $(b_{2g})^1(e_{g})^2(b_{1g})^0(a_{1g})^1$          & 3927.8                                    \\
		${}^{5}E_{g}$  & $(b_{2g})^1(e_{g})^1(b_{1g})^1(a_{1g})^1$          & 31788.2                                   \\
		${}^{5}B_{2g}$ & $(b_{2g})^0(e_{g})^2(b_{1g})^1(a_{1g})^1$          & 37376.8                                        \\  \colrule
		\multicolumn{3}{c}{$D_{2h} (\phi = 103^\circ)$}                                                                                                            \\  \colrule 
		Energy term    & Configuration                                      & Energy \\
		${}^{5}A_{g}$  & $(b_{2g})^1(b_{1g})^1(b_{3g})^1(a_{g})^1(a_{g})^0$ & 0.0                                       \\
		${}^{5}A_{g}$  & $(b_{2g})^1(b_{1g})^1(b_{3g})^1(a_{g})^0(a_{g})^1$ & 4306.1                                    \\
		${}^{5}B_{3g}$ & $(b_{2g})^1(b_{1g})^1(b_{3g})^0(a_{g})^1(a_{g})^1$ & 35052.5                                    \\
		${}^{5}B_{1g}$ & $(b_{2g})^1(b_{1g})^0(b_{3g})^1(a_{g})^1(a_{g})^1$ & 37218.1                                    \\
		${}^{5}B_{2g}$ & $(b_{2g})^0(b_{1g})^1(b_{3g})^1(a_{g})^1(a_{g})^1$ & 38665.4                                      
	\end{tabular}
	\end{ruledtabular}
\end{table}

The third column shows the term energies of the quintet states relative to the lowest state ($^5A_{1g}$ for $D_{4h}$ symmetry and $^5A_{g}$ for $D_{2h}$ symmetry), 
whose energy is set to zero. Within the independent-particle approximation, utilizing these term energies and the orbital populations allows one to derive the 
one-electron energies listed in Table~\ref{tab:table32}. These one-electron energies are measured relative to the lowest one-electron state, $b_{2g}$, which is taken as the energy zero.
\begin{table}[b]
\caption{\label{tab:table32} Orbital energies derived from the atomic Ru $4d$ orbitals for the \ce{[RuO6]^{8-}} system in fields of $D_{4h}$ ($\phi=90^\circ$) and $D_{2h}$ (at the experimental angle $\phi$) symmetry. These energies are extracted from the data in Table~\ref{tab:table3} by configuration-population analysis. The first column shows the orbital and its atomic origin. The second column provides the energies in~$\text{cm}^{-1}$ relative to the lowest orbital energy of the respective point group.}
	\begin{ruledtabular}
	\begin{tabular}{ccccc}
		\multicolumn{2}{c}{$D_{4h}$}                                                                                                                        \\  \colrule
Orbital                 &  Energy \\
 $(a_{1g}):d_{z^2}$      & 37376.8                  \\
 $(b_{1g}):d_{x^2-y^2}$  & 33448.9                  \\
 $(e_{g}):d_{xz},d_{yz}$ & 1660.7                   \\
 $(b_{2g}):d_{xy}$       & 0                        \\  \colrule
		\multicolumn{2}{c}{$D_{2h} (\phi = 103^\circ)$}                                                                                                            \\  \colrule
 Orbital                 &  Energy \\
 $(a_{g}):d_{z^2}$       & 38665.4                  \\
 $(a_{g}):d_{x^2-y^2}$   & 34359.3                  \\
 $(b_{3g}):d_{yz}$       & 3612.8                   \\
 $(b_{1g}):d_{xy}$       & 1447.3                   \\
 $(b_{2g}):d_{xz}$       & 0.0                     
	\end{tabular}
	\end{ruledtabular}
\end{table}

The data in Table~\ref{tab:table32} also permit the derivation of crystal-field splitting parameters for a spherically averaged $O_h$ symmetry environment. From the $D_{4h}$ data, averaging the energies of the two lowest orbitals ($e_{g}$, $b_{2g}$) and the two highest orbitals ($a_{1g}$ and $b_{1g}$ orbitals) yields $\Delta = 32\times10^3\,\text{cm}^{-1}$. From the $D_{2h}$ data, averaging the two highest orbitals ($2a_{g}$) and the three lowest orbitals ($b_{3g}$, $b_{1g}$, and $b_{2g}$) gives $\Delta = 35\times10^3\,\text{cm}^{-1}$. These values can serve as estimates of the $10Dq$ crystal-field splitting parameter. For the actual \ce{RuO2} crystal, this parameter is reported as $10Dq = 21\times10^3\,\text{cm}^{-1}$ in Ref.~\cite{occhialini2021local}. It should be noted, however, that our results were obtained from \textit{ab initio} calculations on the $[\ce{RuO6}]^{8-}$ model, whereas the literature value was derived by comparing experimental data for the \ce{RuO2} crystal with the corresponding parameters of the model used in that study.

\begin{figure}[b]
  \centering
  \includegraphics[width=0.98\columnwidth]{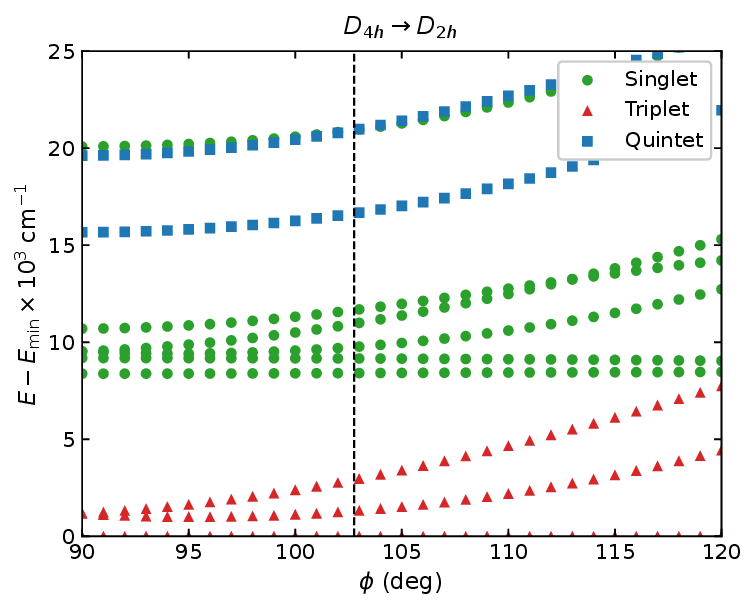}
\caption{\label{fig:en_phi} Term energies for the \ce{[RuO6]^{8-}} system plotted as a function of the angle $\phi$ between the diagonals of the distorted octahedron in the $xy$ equatorial plane, showing the transition from $D_{4h}$ (at $\phi=90^\circ$) to $D_{2h}$ symmetry. The vertical dashed line indicates the angle corresponding to the experimental geometry. The energies are given in~$\text{cm}^{-1}$ relative to the lowest-energy state ($E_{\text{min}}$).}
\end{figure}

In Fig.~\ref{fig:en_phi}, we present the term energies as functions of the angle $\phi$. Across the entire range of angles, the triplet remains the ground state of the system, which is consistent with the results of previous studies~\cite{occhialini2021local,berlijn2017itinerant}. This  result deviates from the predictions of the electrostatic approximation (the point-charge $\ce{Ru^{4+} + 6Q}$ model discussed in Sec.~\ref{sec:ru6q}) for $D_{2h}$ symmetry in the region $|q| > 7$, where the ground state is a singlet.

We note that the data for Fig.~\ref{fig:cas_mo_symm} were obtained without accounting for dynamical correlations, whereas the data presented in Table~\ref{tab:table3} include dynamic electron correlation within the NEVPT2 framework. The data in Table~\ref{tab:table32} show that despite the apparent crossing of the $d_{xz}$ and $d_{xy}$ levels observed in Fig.~\ref{fig:cas_mo_symm}, the inclusion of dynamic electron correlation within NEVPT2 clearly lifts near-degeneracies and separates 
these states, maintaining an energy gap of 1447~$\text{cm}^{-1}$. It is also worth emphasizing that separating the wavefunctions into ``pure'' $d_{xz}$ and $d_{yz}$ 
orbitals is strictly valid only within the independent-particle approximation. In a general multielectron context, the corresponding states contain a superposition of the form $\alpha d_{xz} \pm \beta d_{yz}$.

\subsection{\label{sec:ruo6_ctep} The \ce{[RuO6]{@}CTEP} cluster}
The most accurate model of the \ce{RuO2} crystal examined in this work is a single-\ce{Ru} cluster constructed using the CTEP technique. In this approach, a central \ce{Ru} atom and its six octahedrally coordinated \ce{O} neighbors are embedded within an array of point charges located at the positions of the external \ce{O} anions, complemented by semilocal pseudopotentials (CTPP) centered at the external \ce{Ru} cations. The methodology for constructing this \ce{[RuO6]{@}CTEP} model is described in detail in Sec.~\ref{sec:cluster}. Ultimately, this cluster model provides a comprehensive description of the local electronic structure of the ruthenium dioxide (\ce{RuO2}) crystal.

\begin{figure*}[t]
  \centering
  \includegraphics[width=0.18\textwidth]{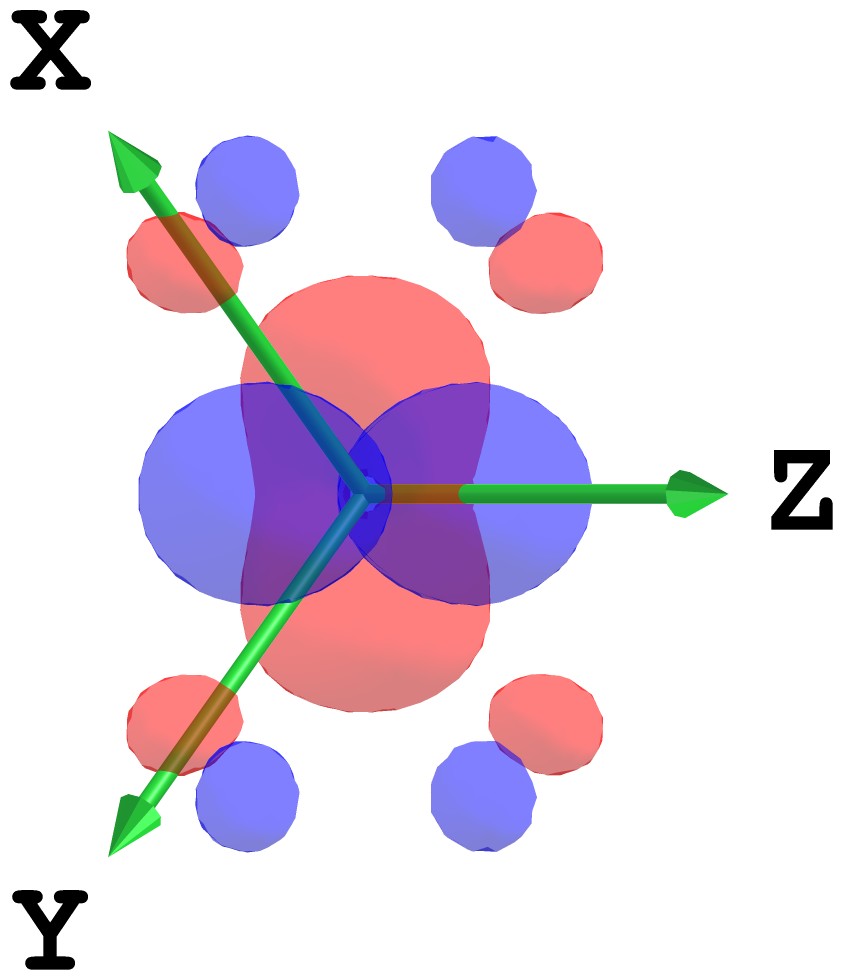}\hfill
  \includegraphics[width=0.18\textwidth]{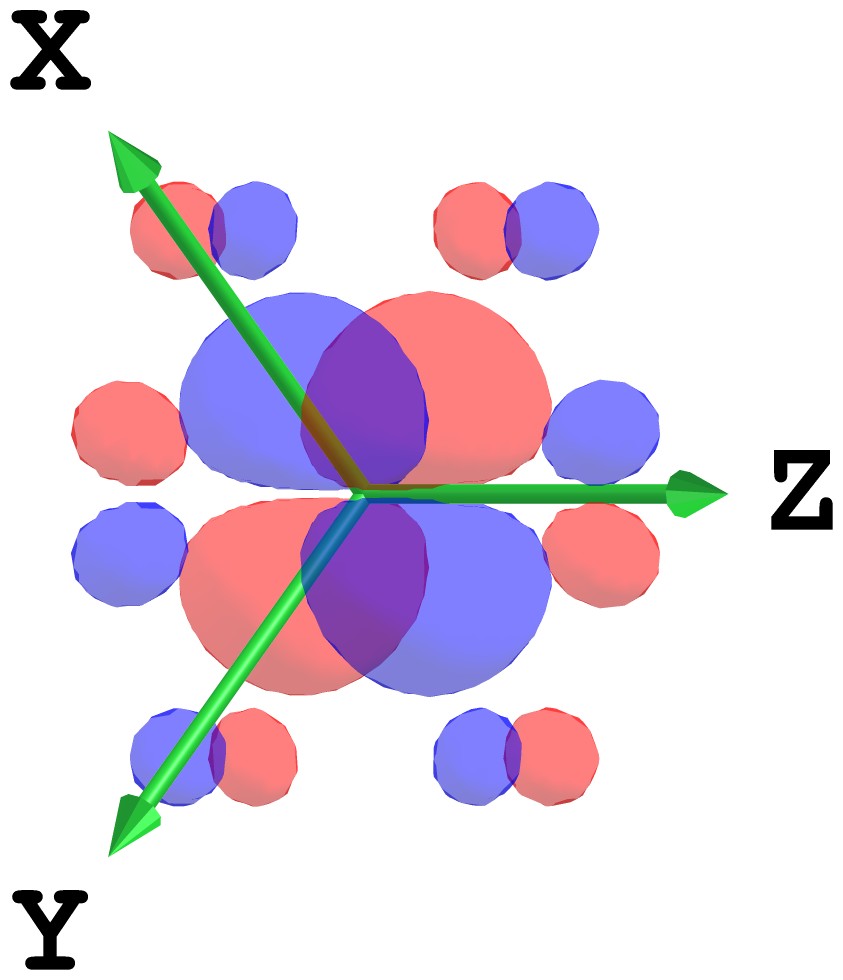}\hfill
  \includegraphics[width=0.18\textwidth]{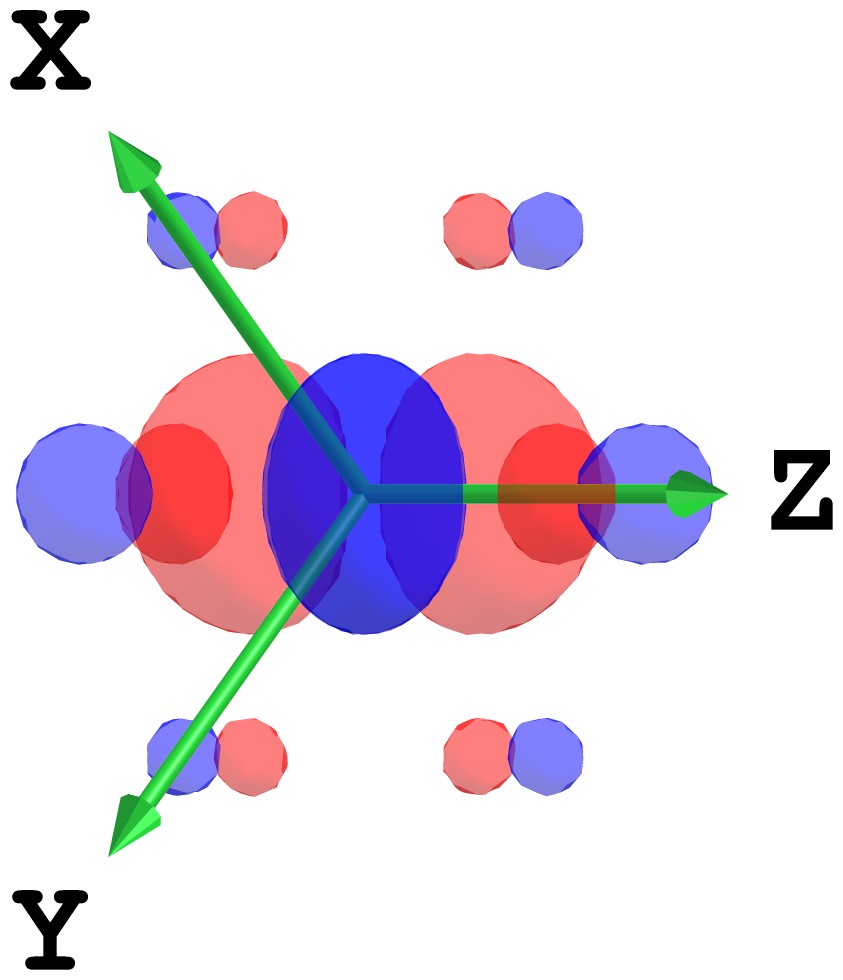}\hfill
  \includegraphics[width=0.18\textwidth]{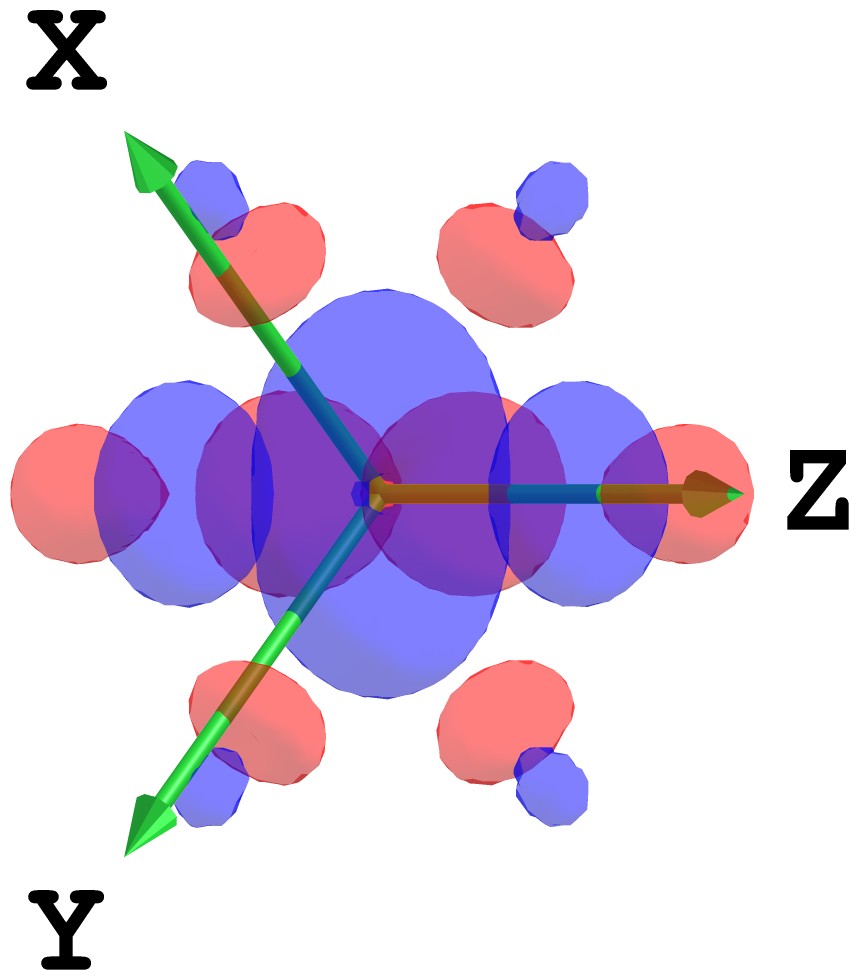}\hfill
  \includegraphics[width=0.18\textwidth]{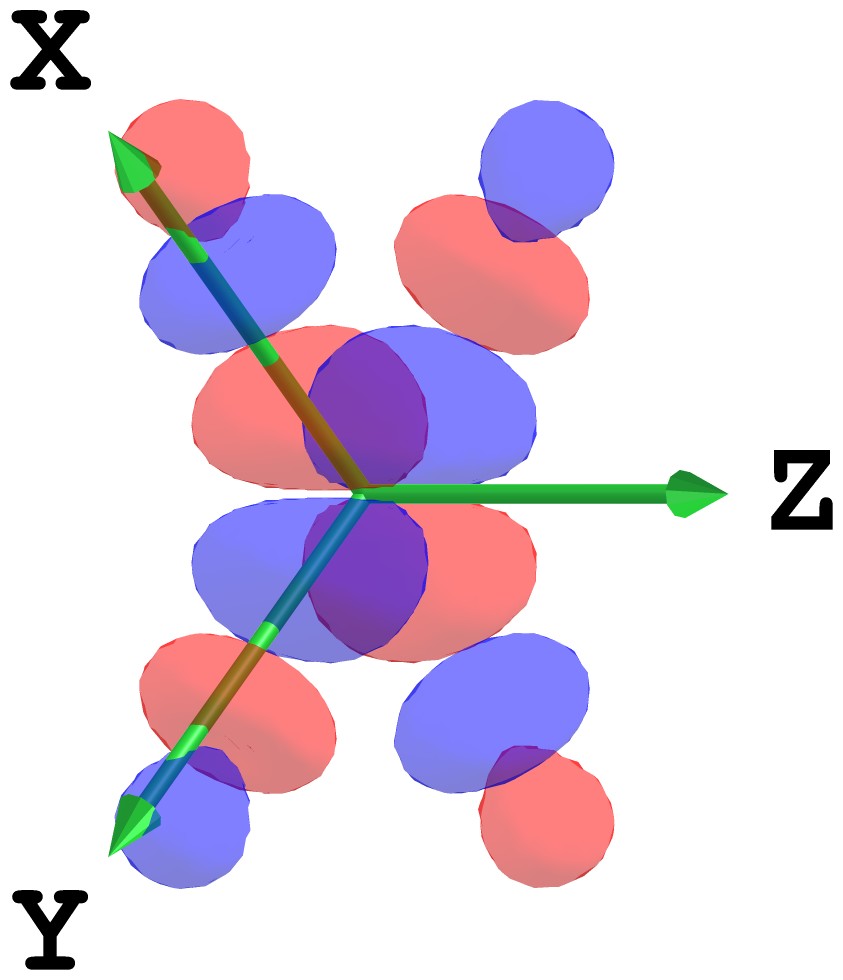}
\caption{\label{fig:mo} Active space orbitals extracted from the SA-CAS(4e,5o) calculation for the  \ce{[RuO6]{@}CTEP} system, associated with the atomic Ru $4d$ orbitals.}
\end{figure*}

The minimal active space capable of adequately describing the local electronic states of the ruthenium ion is SA-CAS(4e,5o), which includes only its $4d$ orbitals. The spatial configuration of the molecular orbitals within this active space is shown in Fig.~\ref{fig:mo}. Although these orbitals retain a typical petal-like shape, their radial and angular densities are noticeably distorted, deviating from the spherical symmetry of a free ion under the influence of the low-symmetry local crystal lattice field.

\begin{figure}[b]
  \centering
  \includegraphics[width=0.98\columnwidth]{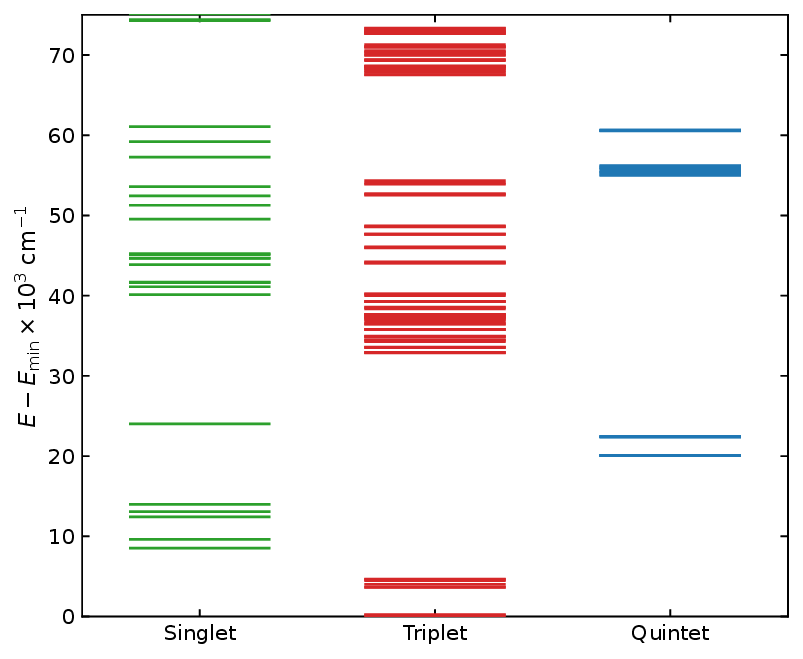}
\caption{\label{fig:soc_en} Term energies for the \ce{RuO6}@CTEP cluster. The data are obtained at the CASSCF+NEVPT2+SOC level of theory. The energies are given in~$\text{cm}^{-1}$ relative to the ground state ($E_{\text{min}}$).}
\end{figure}

The energy spectrum of the \ce{[RuO6]{@}CTEP} cluster was calculated at the CASSCF+NEVPT2+SOC level of theory. The technical details of these calculations are presented in Sec.~\ref{sec:ma_calc}. The resulting energy levels are shown in Fig.~\ref{fig:soc_en}. As seen from the figure, a triplet state remains the ground state of the cluster. Furthermore, it is important to emphasize that within the range from $0$ to $25\times10^3\,\text{cm}^{-1}$, the overall pattern of the energy spectrum is qualitatively closer to the splitting pattern of a $D_{4h}$ symmetry field rather than the $D_{2h}$ symmetry shown in Fig.~\ref{fig:nevpt2_en}. In particular, the diagram for $D_{4h}$ symmetry in Fig.~\ref{fig:nevpt2_en} within the region $7 < |q| < 8$ qualitatively reproduces the energy spectrum pattern of the \ce{[RuO6]{@}CTEP} cluster well. Accordingly, within an effective independent-particle framework, the electronic structure of the \ce{[RuO6]{@}CTEP} cluster closely mimics $D_{4h}$ symmetry. To strengthen this hypothesis, we analyze the energy spectrum in greater detail.

The many-electron ground state of the \ce{[RuO6]{@}CTEP} cluster originates from the ${}^{3}T_{1g}$ term of $O_h$ symmetry. This term splits into the $A_{2g}$ and $E_{g}$ states under $D_{4h}$ symmetry, and further decomposes into the $B_{1g}$, $B_{2g}$, and $B_{3g}$ states under the true $D_{2h}$ local symmetry of the \ce{[RuO6]{@}CTEP} cluster. The calculated energy levels are listed in Table~\ref{tab:table4}. The first column shows the spin multiplicity of the states. Columns $2$--$4$ list the corresponding irreducible representations for $O_h$, $D_{4h}$, and $D_{2h}$ symmetries, respectively. The fifth column indexes the energy levels in ascending order, and the final column provides the relative energies measured from the ground state.

The $9$ microstates corresponding to the $^3T_{1g}$ term in $O_h$ symmetry must partition into two sets under $D_{4h}$ symmetry: $3$ states belonging to the $^3A_{2g}$ term and $6$ states belonging to the $^3E_{g}$ term. Under $D_{2h}$ symmetry, these must partition into three sets containing $3$ states for each of the $^3B_{1g}$, $^3B_{2g}$, and $^3B_{3g}$ terms. Our calculations show that the energy levels can be naturally grouped into two distinct sets ($3+6$) corresponding to $D_{4h}$ symmetry. The manifestation of the symmetry reduction from $D_{4h}$ to $D_{2h}$ (i.e., the further lifting of degeneracy within the $3+6$ groupings into a $3+3+3$ arrangement) is remarkably weak.

Consequently, while the formal local symmetry of the \ce{Ru} site in the \ce{RuO2} crystal is $D_{2h}$, which theoretically lifts all spatial degeneracy and results in $9$ orbitally non-degenerate levels, an analysis of the ground $^3T_{1g}$ term splitting reveals a different effective picture. We can see that in Table~\ref{tab:table4} for $^3T_{1g}$ term, that $3$ of $9$ microstates are strongly quasi-degenrate. This degeneracy corresponds more typical for $D_{4h}$ than $D_{2h}$ symmetry.

Based on the calculated spectrum, we conclude that the highest six levels (states $4$--$9$ of the $^3T_{1g}$ manifold in Table~\ref{tab:table4}) are likewise quasi-degenerate due to the very small energy gaps separating them. This behavior clearly indicates that the orthorhombic $D_{2h}$ component of the crystal field in \ce{RuO2} acts merely as a weak perturbation relative to the dominant tetragonal $D_{4h}$ distortion.

On the one hand, an analysis of the spectrum in Fig.~\ref{fig:soc_en} mapped onto a independent-particle approximation is valuable because such frameworks are conventionally implemented to model the magnetic structure of altermagnets. On the other hand, an effective independent-particle analysis can be carried out, similar to the one employed in Section~\ref{sec:ruo6}. As before, the wavefunction of the quintet many-electron terms remains nearly one-determinant within the active space (with weights \(\ge 99\%\)). Nevertheless, all the previously discussed limitations remain valid (this framework neglects the multiconfigurational nature of the investigated many-electron states). Furthermore, determining the symmetry of the quintets and, consequently, the symmetry of the constituent orbitals becomes impossible due to the specific nature of the cluster model.

Despite these limitations, such an analysis remains essential because it enables a qualitative assessment of the contributions from dynamic correlation and SOC. In contrast to the case discussed in Section~\ref{sec:ruo6}, where these corrections lifted the degeneracy and separated the states, here they conversely lead to the quasi-degeneracy of the energy levels. Specifically, the energy separation between the third and fourth terms does not exceed 200--300$\,\text{cm}^{-1}$. This value is comparable to the thermal energy $k_{B}T$ at room temperature, suggesting that the states are quasi-degenerate in a real crystal.
This observation strongly suggests that the electronic states are effectively governed by the $D_{4h}$ symmetry rather than the local $D_{2h}$ symmetry.

We would also like to note that we performed DFT calculations (without the Hubbard potential $U$) for the \ce{[RuO6]}@CTEP cluster. In contrast to the multiconfigurational results, DFT does not reveal the strong quasi-degeneracy of the third and fourth terms. Instead, it yields a splitting of approximately $1000\,\mathrm{cm}^{-1}$. This can be regarded as evidence that DFT fails to provide a qualitatively correct description of the local electronic structure of the \ce{RuO_{2}} crystal.

\begin{table*}
\caption{\label{tab:table4} Term energies of the lowest states for the \ce{[RuO6]} CTEP cluster. The first column shows the multiplicity of the terms. The energy term symbol column lists the labels assigned according to the evolution of states under successive symmetry lowering ($O_h \rightarrow D_{4h} \rightarrow D_{2h}$). The `State No.' column gives the state numbers. The last column provides the term energies in~$\text{cm}^{-1}$ relative to the lowest-energy state.}
        \begin{ruledtabular}
	\begin{tabular}{ccccccc}
		$2S+1$ & \multicolumn{3}{c}{Energy term symbol}                                        & State No. & Energy difference  & Relative energy  \\
		& $O_h$  & $D_{4h}$                  & $D_{2h}$                                &       &                                      &                                          \\  \colrule
		3                    & $T_{1g}$ & $A_{2g} \oplus E_{g}$  & $B_{1g} \oplus B_{2g} \oplus B_{3g}$ & 1     & 0                                    & 0                                        \\
		&        &                        &                                      & 2     & 200                                  & 200                                      \\
		&        &                        &                                      & 3     & 25                                   & 225                                      \\
		&        &                        &                                      & 4     & 3413                                 & 3638                                     \\
		&        &                        &                                      & 5     & 18                                   & 3656                                     \\
		&        &                        &                                      & 6     & 370                                  & 4026                                     \\
		&        &                        &                                      & 7     & 445                                  & 4471                                     \\
		&        &                        &                                      & 8     & 177                                  & 4648                                     \\
		&        &                        &                                      & 9     & 28                                   & 4676                                     \\  \colrule
		1                    & $E_g$  & $A_{1g} \oplus B_{1g}$ & $A_{g} \oplus B_{1g}$                & 1--2     & -                                    & 9076                                     \\
		1                    & $T_{2g}$ & $B_{2g} \oplus E_{g}$  & $A_{g} \oplus B_{2g} \oplus B_{3g}$  & 1--3     & -                                    & 13162                                    \\
		5                    & $E_g$  & $A_{1g} \oplus B_{1g}$ & $A_{g} \oplus B_{1g}$                & 1--10    & -                                    & 22410                                    \\
		1                    & $A_{1g}$ & $A_{1g}$                 & $A_g$                                & 1     & -                                    & 24011                                   
    \end{tabular}
	\end{ruledtabular}
\end{table*}

\section{Discussion}
The observation of altermagnetism in ruthenium dioxide (\ce{RuO2}) is currently a subject of intense discussion. Several research groups have reported evidence supporting the presence of altermagnetic order~\cite{berlijn2017itinerant,Zhu2019PhysRevLett.122.017202,Feng2022Nature,Bose2022,Bai2022PhysRevLett.128.197202,Karube2022PhysRevLett.129.137201,He2025_NatCommun,Akashdeep2026}, while other experimental studies point to an apparent absence of altermagnetic signatures~\cite{Hiraishi2024PhysRevLett.132.166702,Kessler2024,Song2025}. A number of theoretical works based on independent-particle approximations not only predict altermagnetism in \ce{RuO2} crystals but also model its resulting properties. Crucially, the existence of altermagnetism relies on the distinct contrast between the global and local symmetries of the ruthenium atoms.
While the global crystal lattice point group belongs to the tetragonal $D_{4h}$ system, the exact local site symmetry of the \ce{Ru} atoms, surrounded by a distorted oxygen octahedron, is reduced to the orthorhombic $D_{2h}$ point group.

In this work, we theoretically investigated the electronic structure of the $\ce{RuO2}$ crystal. Our multielectron cluster calculations directly demonstrate that the effective local crystal field is symmetry-wise much closer to the higher-symmetry $D_{4h}$ group than to the formal orthorhombic $D_{2h}$ point group. Notably, it is precisely the orthorhombic $D_{2h}$ component of the local environment that is theoretically expected to enable the emergence of a local $xy$ quadrupole order on the two ruthenium magnetic sublattices. In existing independent-particle theoretical models~\cite{roig2024minimal}, this point-group symmetry reduction serves as the primary mechanism driving altermagnetic spin splitting. However, our ground-state calculations reveal that the relevant $4d$ orbitals in a stoichiometric bulk crystal retain a strong quasi-degeneracy. Consequently, the local quadrupole order that would otherwise arise from the orthorhombic distortion is heavily suppressed, rendering it insufficiently large to manifest observable altermagnetism.

This perspective provides a plausible explanation for recent experimental developments. While some bulk measurements see no altermagnetic signatures, compelling evidence for altermagnetism has been reported in \ce{RuO2}(101) thin films synthesized via fully epitaxial growth on sapphire substrates~\cite{He2025_NatCommun,Akashdeep2026}. Based on this dichotomy, we hypothesize that the macroscopic electronic and magnetic properties of \ce{RuO2} are highly sensitive to sample preparation conditions and strain engineering. In a pristine, stoichiometric bulk single crystal, the quasi-degeneracy of the $4d$ orbitals acts to minimize the local quadrupole order, thereby preventing the stabilization of a robust altermagnetic phase. In contrast, in thin films, where biaxial compression or tension is exerted by the substrate, or in the presence of structural point defects, the local orthorhombic distortion can be significantly enhanced. Such external perturbations lift the orbital quasi-degeneracy, stabilizing the local quadrupole order and ultimately enabling the distinct altermagnetic properties observed experimentally. Future work modeling these surface and strain effects via customized cluster models is currently underway.

\section{Conclusion}
We have theoretically investigated the local electronic structure of the \ce{RuO2} crystal using a multielectron \ce{[RuO6]{@}CTEP} cluster model calculated at the CASSCF+NEVPT2+SOC level of theory. Our main findings demonstrate that while the formal local site symmetry of the ruthenium ion is orthorhombic ($D_{2h}$), the calculated energy spectrum and orbital splitting pattern behave much closer to a higher tetragonal ($D_{4h}$) symmetry. Crucially, this means that the effective implementation of the independent-particle approximations, which are conventionally used to model the altermagnetic spin splitting, overestimates the local symmetry reduction. In reality, our multielectron approach shows that this effective higher symmetry preserves a strong quasi-degeneracy among the relevant $4d$ orbitals within the stoichiometric bulk crystal. 

This persistent quasi-degeneracy suppresses the formation of the local $xy$ quadrupole order on the ruthenium sublattices that would otherwise be driven by the orthorhombic distortion. Because this quadrupole order is the primary mechanism responsible for altermagnetic spin splitting in independent-particle models, its suppression explains why experimental bulk measurements observe an absence of altermagnetism \cite{Hiraishi2024PhysRevLett.132.166702,Kessler2024,Song2025}. Finally, we propose that external perturbations, such as epitaxial strain in thin films or local structural point defects, act to break this orbital quasi-degeneracy, thereby stabilizing the quadrupole order and enabling the robust altermagnetic properties observed in recent thin-film experiments \cite{He2025_NatCommun,Akashdeep2026}.

\begin{acknowledgments}
The authors thank A.V.~Titov for reading the manuscript and for his help, and A.V.~Oleynichenko for valuable discussions.

This work was carried out using computing resources of the federal
collective usage center, Complex for Simulation and Data Processing for Megascience
Facilities at National Research Center ''Kurchatov Institute'', http://ckp.nrcki.ru/.

The work was supported by the Russian Science Foundation and the St. Petersburg
Science Foundation (grant No. 25-23-20046).

\end{acknowledgments}

\appendix

\section{\label{sec:lc-ecp}Large-core pseudopotentials for effective state of atom in crystal}
\label{lc-ecp}
In the CTEP model, the NCE is constructed from pseudoatoms represented by large-core pseudopotentials (lc-PPs) with additional fractional charges and truncated basis sets. While these lc-PPs are not aimed to yield precise description of the corresponding pseudoatom properties in a wide range of changes, their influence on the main cluster area of the embedding cluster should be as close as possible to that of corresponding atoms of the original crystal. The large-core approach serves two main goals: a) to decrease computational cost of the CTEP model, and b) to avoid unphysical core relaxation when fractional charges are introduced. However, it is widely known that lc-PPs exhibit considerable less transferability than small- or medium-core ones. So, if lc-PPs are made for one effective state and applied to a different one, significant errors are likely to emerge. 

In the previous CTEP papers \cite{lomachuk2020compound,maltsev2021compound,shakhova2022compound,oleynichenko2024compound,maltsev2025electronic}, we introduced and utilized a conception of CTPPs. In this framework, the lc-PPs were initially made by shape-consistent approach for some fractional averaged state of an isolated atom and then ``tuned'' it for the effective crystal state. The PP-construction method was same as \GRECP ~one \cite{titov1999generalized} except that semilocal-only form was used as there is no outer core in lc-PP. The effective atomic orbital occupations were selected from either Mulliken or atom-in-compound (AiC) \cite{titov2014concept} population analysis. To tune the resulting lc-PPs to match an effective state in a given compound, a parameterization and optimization procedure was performed.
Specifically, the lc-PPs were parameterized (for instance, by introducing an initially zero correction term), and selected parameters were optimized to minimize the RMS energy gradient of the main crystal region, where all cations of a selected type were represented as lc-PPs instead of the original \GRECP ~(small-core pseudopotentials). While this scheme yielded satisfactory results, it suffered from a major drawback: the initial
lc-PPs often yielded excessively large starting RMS gradients. This shifted the CTPP optimizing process from intended small ``tuning'' to significant PP rebuilding. The most likely reason for such discrepancies is poor correspondence between the isolated atomic state and its effective state within the crystal.

In the present paper, we exploit a new scheme to construct the initial lc-PP approximation for an effective crystal state. This method is based on simulating crystal electronic and nuclear environment in a spherically averaged atomic calculation (spherical symmetry is required for a shape-consistent PP building). First, a density matrix in atomic Gaussian
function basis $\left\{ \chi_{\mu}\right\} $ is obtained from a CRYSTAL
\cite{dovesi2018crystal17,erba2022crystal23,crystal23}
properties calculation for a cluster containing the atom of interest and all its neighbors. The total radial density is then evaluated by numeric integration using selected radial grid and a Lebedev angular grid:
\begin{eqnarray}
D(r)
&=&
4\pi r^{2}\sum_{k}w_{k}\sum_{\mu\nu}P_{\mu\nu}\chi_{\mu}(r,\Omega_{k})\chi_{\nu}(r,\Omega_{k})
\,,
\end{eqnarray}
where $\Omega_{k}=\left(\theta_{k},\phi_{k}\right)$ and $w_{k}$ denote the Lebedev grid coordinates and weights, respectively, $P_{\mu\nu}$ is the total density matrix in $\left\{ \chi_{\mu}\right\} $ basis. The radial grid is constructed by smoothly sewing three distinct regions: a logarithmic grid in the central region, a linear grid in the region of the neighboring atoms, and another logarithmic grid at large $r$ values. The $l$-components of radial density up to $l=15$ are generated by projecting the basis functions onto corresponding spherical harmonics:
\begin{eqnarray}
A_{\mu,lm}(r)=\sum_{k}w_{k}\chi_{\mu}(r,\Omega_{k})Y_{lm}(\Omega_{k})
\,,
\\
D_{l}(r)=4\pi r^{2}\sum_{m=-l}^{l}\sum_{\mu\nu}P_{\mu\nu}A_{\mu,lm}(r)A_{\nu,lm}(r)
\,.
\end{eqnarray}
The remainder $D(r)-\sum_{l}D_{l}(r)$ is treated as $D_{\max l+1}(r)$. 
The resulting $D_{l}(r)$ densities are subsequently used in an atomic HF/DFT calculation in a spline basis with the original small-core pseudopotential. The spline knot grid is constructed using a similar three-part approach to that used for evaluating $D_{l}(r)$, but with different parameters. To ensure consistency with the original calculation, the same exchange-correlation potential is employed. As spherical symmetry is required, only one radial function is used for each $\left(n,l\right)$ pair, however, fractional occupation numbers are allowed to simulate ``fraction of a closed shell''.

The effective crystal environment is constructed from three components. First, each neighbor atomic core is modeled by a charged sphere potential (Watson sphere) with radius equal to interatomic distance and charge equal to corresponding atomic number (or the number of valence electrons if pseudopotentials are also used for the neighboring atoms). Second, the radial densities obtained on previous step are used to generate ``environment quasi-orbitals''. At each HF/DFT iteration, the total density of atomic orbitals, $\rho_{l}(r)=\sum_{n}n_{nl}|\psi_{nl}|^{2}(r)$, for each $l$ is subtracted from corresponding $D_{l}(r)$. The square root of the remainder is treated as quasi-orbital with corresponding occupation number:
\begin{eqnarray}
\tilde{\rho}_{l}(r)
&=&
\max\left(0,D_{l}(r)-\rho_{l}(r)\right)
\,,
\\
\tilde{n}_{l}
&=&
\int\tilde{\rho}_{l}(r)dr
\,,
\\
\tilde{\phi}_{l}(r)
&=&
\sqrt{\frac{\tilde{\rho}_{l}(r)}{\tilde{n}_{l}}}
\,.
\end{eqnarray}

The resulting $\tilde{\phi}_{l}(r)$ and $\tilde{n}_{l}$ for each $l$ value are included in evaluating two-electron Coulomb and exchange matrices and exchange-correlation potential in the subsequent HF/DFT iteration.
The third part of the environment consists of a set of particular parameterized
``localizing potentials'' $L_{l}(r)$ for each $l$ value which
effectively simulate the Pauli repulsion originating from the electronic environment.
As $L_{l}(r)$ and the atomic occupation numbers $n_{nl}$ are initially
unknown, a two-level fitting procedure is performed. At the ``low
level'' for a given set of $L_{l}(r)$, the occupation numbers $n_{nl}$
are iteratively fitted so that the atomic densities $\rho_{l}(r)$ match
the original $D_{l}(r)$ as closely as possible within  a specified region around the center. At the ``high level'' the parameters of $L_{l}(r)$
are optimized with criterion of minimizing this low-level matching
error. The maximum $l$ value for the occupied atomic orbitals $\psi_{nl}$
defines the local part of the subsequently constructed PP and is selected
by both physical intuition and trial-and-error method. The quasi-orbitals,
however, are constructed and utilized for all $l$-values  present in the input $D_{l}(r)$
set.

Finally, the atomic calculation in the spline basis with the optimized occupation numbers and environment parameters is used for a PP building following the scheme described in \cite{titov1999generalized}. A scalar-only semilocal PP is constructed, though relativistic effects are implicitly included by employing a relativistic small-core PP in the initial calculation. When deployed in crystal calculations, the resulting lc-PPs yield significantly smaller initial RMS gradients than those constructed from an isolated atomic state. Although these initial RMS values are still not perfectly zero and subsequent tuning remains necessary to obtain fully optimized CTPPs, the scale of this tuning is successfully reduced to a minor correction.


\section{\label{sec:a_svd}Electrostatic SVD analysis of the cluster}
\label{svd}
One of the most critical and computationally demanding phases in constructing an embedded crystal cluster model is the optimization of the embedding point charges. This procedure entails selecting both the spatial coordinates and the magnitudes of the point charges. Typically, these charges are positioned at the atomic sites immediately surrounding the active region of the cluster. Their initial values are assigned based on formal oxidation states, electronic population analyses, and the requirement of global charge neutrality (or a specified net charge for the cluster). The embedding optimization then iteratively refines these values based on two primary criteria: first, minimizing the nuclear forces (energy gradients) acting on the atoms within the active quantum cluster; second, ensuring the physical plausibility of the refined charges by maintaining their proximity to formal oxidation states. For small clusters, the number of point charges ($n_q$) can scale to several dozen, while for larger models, it can easily exceed one hundred. This set of point charges defines an $n_q$-dimensional configuration space, where selecting an optimal basis of linear combinations of charges is of paramount importance.

If the active subsystem of the cluster contains $n_a$ atoms, the primary objective of the charge optimization is to vanish or minimize the $3n_a$ energy gradients (three spatial components per atom). These gradients span a $3n_a$-dimensional space. Generally, to achieve a robust reduction of these gradients, the dimensionality of the point-charge space must exceed that of the gradient space ($n_q > 3n_a$). Our immediate task is to identify a well-behaved subspace within the charge space to solve for the charge vector that minimizes the structural gradients. To achieve this, we employ an electrostatic approximation, assuming that the underlying crystal symmetry dictates the mapping between the charge configuration space and the gradient response space.

Let the atoms within the active part of the cluster be characterized by nuclear charges $\{Z_\alpha\}_{\alpha = 1}^{n_a}$ and coordinate vectors $\{\zhr_\alpha\}$, which we denote as $\zhr_\alpha = \{r_{i\alpha}\}_{i=1,2,3}$. Furthermore, let the embedding point charges $\{q_j\}_{j=1}^{n_q}$ form a vector $\zhq$ and be located at coordinates $\{\zhrho_j\}$, denoted as $\zhrho_j = \{\rho_{ij}\}_{i=1,2,3}$. 

We isolate the part of the cluster Hamiltonian that explicitly depends on the embedding point charges:
\begin{eqnarray}
H
&=&
H_0 + H_q(\zhq)
\,,
\\
H_q(\zhq)
&=&
\sum_{\alpha=1}^{n_a} \sum_{j=1}^{n_q} \frac{Z_{\alpha} q_j}{|\zhr_{\alpha} - \zhrho_j|}
\,,
\end{eqnarray}
where $H_0$ encompasses all terms independent of $\zhq$.

To linearize the response, we evaluate the response matrix of mixed derivatives of the total cluster energy ($E$) with respect to nuclear displacements and charge modifications:
\begin{eqnarray}
A_{i\alpha,j}
&=&
\frac{\partial^2 E}{\partial r_{i\alpha} \partial q_j}
\,.
\end{eqnarray}
This matrix maps variations in the embedding charges directly to the resulting variations in the nuclear forces. 

A reliable zeroth-order approximation for this response matrix can be derived purely within the electrostatic limit:
\begin{eqnarray}
A_{i\alpha,j}^{(0)}
&=&
\frac{\partial^2 H_q}{\partial r_{i\alpha} \partial q_j} = \frac{Z_{\alpha}}{|\zhr_{\alpha} - \zhrho_j|^3} (\rho_{ij} - r_{i\alpha})
\,.
\end{eqnarray}.
In this limit, the $i$-th component of the force gradient acting on atom $\alpha$ simplifies to:
\begin{eqnarray}
g_{\alpha i}^{(0)}
&=&
\frac{\partial H_q}{\partial r_{i\alpha}}
\,.
\end{eqnarray}
The electrostatic matrix $A^{(0)}$ thus establishes the approximate mapping from the charge space to the gradient space. 

To analyze its spectral properties, we perform a singular value decomposition (SVD) on $A^{(0)}$:
\begin{eqnarray}
A^{(0)}
&=&
U^{(0)} S^{(0)} V^{(0)\dagger}
\,,
\end{eqnarray}
where $U^{(0)}$ is a $3n_a \times 3n_a$ unitary matrix, $V^{(0)}$ is an $n_q \times n_q$ unitary matrix, and $S^{(0)}$ is a $3n_a \times n_q$ rectangular diagonal matrix. The matrices $U^{(0)}$ and $V^{(0)}$ can be expressed in terms of their respective column eigenvectors:
\begin{eqnarray}
V^{(0)}
&=&
(\zhv_1, \dots, \zhv_{n_q}), \quad \zhv_j^T \zhv_{j'} = \delta_{jj'}
\,,
\\
U^{(0)}
&=&
(\zhu_1, \dots, \zhu_{3n_a}), \quad \zhu_i^T \zhu_{i'} = \delta_{ii'}
\,.
\end{eqnarray}
The diagonal elements of $S^{(0)}$ represent the singular values, arranged in descending order ($s_i \ge s_{i+1}$):
\begin{eqnarray}
S_{ij}^{(0)}
&=&
\delta_{ij} s_i, \quad \text{where } s_i \ge 0
\,.
\end{eqnarray}
Given that $n_q > 3n_a$, the rectangular matrix $A^{(0)}$ can be expanded as an outer product restricted to the non-zero singular spectrum:
\begin{eqnarray}
A^{(0)}
&=&
\sum_{i=1}^{3n_a} s_i \, \zhu_i \otimes \zhv_i^{\dagger}
\,.
\end{eqnarray}
Note that the rightmost null-space vectors $\{\zhv_i\}_{i > 3n_a}$ do not contribute to this expansion.

Within this electrostatic framework, we can utilize the pseudoinverse of the response matrix to identify the target charge vector that completely cancels out the initial nuclear gradients. The total gradient vector of the cluster can be linearized via a first-order Taylor expansion:
\begin{eqnarray}
g_{i\alpha} \approx g_{i\alpha}^{(0)} + \sum_j \frac{\partial g_{i\alpha}^{(0)}}{\partial q_j} \delta q_j
&=&
g_{i\alpha}^{(0)} + A_{i\alpha,j}^{(0)} \delta q_j
\,,
\end{eqnarray}
where $\zhg^{(0)}$ represents the gradient vector evaluated at the initial trial charge distribution $\zhq^{(0)}$. Imposing the root condition $\zhg = 0$ yields the required charge correction vector:
\begin{eqnarray}
\delta \zhq \approx -(A^{(0)})^{-1} \zhg^{(0)}
\,.
\end{eqnarray}

By virtue of the Hellmann--Feynman theorem, vanishing the structural gradients in the electrostatic approximation (which corresponds to nullifying the net electrostatic field at the nuclei) directly drives the minimization of the true quantum mechanical gradients. Consequently, this expression serves as the foundation for an iterative optimization scheme to find the optimal charge corrections $\delta q_j$. Multiple iterations are necessary because the underlying electronic density relaxes self-consistently in response to updates in $\delta \zhq$.

To efficiently update the charge vector while avoiding full matrix recomputations, we implement a quasi-Newton approach akin to Broyden's method \cite{PULAY1980393}. The updates for the charge vector $\zhq^{(k)}$ are conveniently parameterized using the displacement vectors $\bm{\kappa}^{(k)}$ (with $\bm{\kappa}^{(1)} = 0$):
\begin{eqnarray}
\zhq^{(k+1)}
&=&
\zhq^{(k)} + V^{(0)} \bm{\kappa}^{(k+1)}
\,.
\end{eqnarray}
By introducing an auxiliary matrix $B^{(0)} = U^{(0)} S^{(0)}$, the recursive updates for the vectors $\bm{\kappa}^{(k)}$, $\zhg^{(k)}$, and the quasi-Newton updating matrix $B^{(k)}$ take the following form:
\begin{eqnarray}
\bm{\kappa}^{(k+1)}
&=&
-\lambda \, B^{(k-1)} \zhg^{(k)}
\,,
\\
\zhg^{(k+1)}
&=&
\zhg(\zhq^{(k+1)})
\,,
\\
B^{(k+1)}
&=&\nonumber
B^{(k)}
\\
&&\hspace{-10pt}
+ \bigl(\zhg^{(k+1)} + (\lambda - 1)\zhg^{(k)}\bigr) \otimes \frac{\bm{\kappa}^{(k+1)} }{\|\bm{\kappa}^{(k+1)}\|^2}
\,.
\end{eqnarray}.
Here, $\lambda$ ($0 < \lambda \le 1$) acts as a scalar damping parameter, and the function $\zhg(\zhq)$ denotes the self-consistently computed quantum mechanical gradient vector for a given charge configuration $\zhq$.

In practice, non-electrostatic contributions, such as basis set incompleteness errors and Pulay forces \cite{PULAY1980393}, introduce non-linearities that complicate the optimization path. Nevertheless, the SVD profile of the electrostatic matrix $A^{(0)}$ provides a reliable diagnostic metric for assessing the mathematical conditioning of the constructed CTEP embedding scheme. Unusually small singular values ($s_i \to 0$) act as numerical hazards; when inverting $A^{(0)}$, these modes can generate unphysically large charge oscillations due to weights inversely proportional to $s_i$. Furthermore, physical constraints dictate that the net sum of forces over the active region must remain zero, implying that the optimization path should be restricted to basis vectors that respect the global translational and rotational symmetries of the cluster. If the initial forces cannot be well-spanned by a linear combination of these symmetry-adapted vectors, the charge optimization routine will likely fail to achieve tight gradient convergence.

Ultimately, this SVD approach reaches its maximum efficiency in high-symmetry cluster geometries. In such cases, group theory and symmetry adaptation can dramatically collapse the dimensionality of the active charge subspace from the original $n_q$ down to just 2 or 3 independent degrees of freedom. For lower-symmetry configurations, the optimization subspace remains considerably larger; however, utilizing the SVD-derived basis vectors $\zhv_i$ still significantly accelerates the numerical convergence of the embedding charge field.
\section{\label{sec:sym}Group-theoretical notation and correlation tables}
\label{sym}
Group-theoretical notations are used to describe the symmetry of the one-electron orbitals and many-electron terms. In particular, the notations for the irreducible representations of the $SO(3)$, $O_h$, $D_{4h}$, and $D_{2h}$ symmetry groups are employed
\cite{altmann1994point,Gelessus1995}.

The splitting of the set of $d$ orbitals into subsets forming irreducible representations of the corresponding point group (labeled by Mulliken symbols) is presented in Table~\ref{tab:sym_1}.

\begin{table}[h!]
  \caption{\label{tab:sym_1} Symmetry species of the $d$ orbitals (represented by quadratic Cartesian functions) in the $O_h$, $D_{4h}$, and $D_{2h}$ point groups.}
  \begin{ruledtabular}
    \begin{tabular}{cccc}
      Orbital       & $O_h$    & $D_{4h}$ & $D_{2h}$ \\ \colrule
      $d_{z^2}$     & $E_g$    & $A_{1g}$ & $A_{g}$  \\
      $d_{x^2-y^2}$ & $E_g$    & $B_{1g}$ & $A_{g}$  \\
      $d_{xy}$      & $T_{2g}$ & $B_{2g}$ & $B_{1g}$ \\
      $d_{xz}$      & $T_{2g}$ & $E_g$    & $B_{2g}$ \\
      $d_{yz}$      & $T_{2g}$ & $E_g$    & $B_{3g}$
    \end{tabular}
  \end{ruledtabular}
\end{table}

The symmetry of a many-electron term within the independent-particle approximation can be found from the symmetry of the one-electron orbitals by analyzing the direct product of their irreducible representations. The representation multiplication tables for the corresponding point groups are presented in Table~\ref{tab:sym_2}. In particular, the states listed in Table~\ref{tab:table3} were determined in this way.

\begin{table}[h!]
\caption{\label{tab:sym_2} Direct products of irreducible representations.}
\begin{ruledtabular}
  \begin{tabular}{ccccc}
\multicolumn{5}{c}{$O_h$} \\ \hline
\multicolumn{1}{c|}{} & \multicolumn{2}{c|}{$\mathbf{E_g}$} & \multicolumn{2}{c}{$\mathbf{T_{2g}}$} \\ \hline
\multicolumn{1}{c|}{$\mathbf{E_g}$} & \multicolumn{2}{c|}{$A_{1g} \oplus [A_{2g}] \oplus E_{g}$} & \multicolumn{2}{c}{$T_{1g} \oplus T_{2g}$} \\ \hline
\multicolumn{1}{c|}{$\mathbf{T_{2g}}$} & \multicolumn{2}{c|}{$T_{1g} \oplus T_{2g}$} & \multicolumn{2}{c}{$A_{1g} \oplus E_g \oplus [T_{1g}] \oplus T_{2g}$} \\ \hline
\multicolumn{5}{c}{$D_{4h}$} \\ \hline
\multicolumn{1}{c|}{} & \multicolumn{1}{c|}{$\mathbf{A_{1g}}$} & \multicolumn{1}{c|}{$\mathbf{B_{1g}}$} & \multicolumn{1}{c|}{$\mathbf{B_{2g}}$} & $\mathbf{E_g}$ \\ \hline
\multicolumn{1}{c|}{$\mathbf{A_{1g}}$} & \multicolumn{1}{c|}{$A_{1g}$} & \multicolumn{1}{c|}{$B_{1g}$} & \multicolumn{1}{c|}{$B_{2g}$} & $E_g$ \\ \hline
\multicolumn{1}{c|}{$\mathbf{B_{1g}}$} & \multicolumn{1}{c|}{$B_{1g}$} & \multicolumn{1}{c|}{$A_{1g}$} & \multicolumn{1}{c|}{$A_{2g}$} & $E_g$ \\ \hline
\multicolumn{1}{c|}{$\mathbf{B_{2g}}$} & \multicolumn{1}{c|}{$B_{2g}$} & \multicolumn{1}{c|}{$A_{2g}$} & \multicolumn{1}{c|}{$A_{1g}$} & $E_g$ \\ \hline
\multicolumn{1}{c|}{$\mathbf{E_g}$} & \multicolumn{1}{c|}{$E_g$} & \multicolumn{1}{c|}{$E_g$} & \multicolumn{1}{c|}{$E_g$} & $A_{1g}  \oplus [A_{2g}] \oplus B_{1g} \oplus B_{2g}$ \\ \hline
\multicolumn{5}{c}{$D_{2h}$} \\ \hline
\multicolumn{1}{c|}{} & \multicolumn{1}{c|}{$\mathbf{A_g}$} & \multicolumn{1}{c|}{$\mathbf{B_{1g}}$} & \multicolumn{1}{c|}{$\mathbf{B_{2g}}$} & $\mathbf{B_{3g}}$ \\ \hline
\multicolumn{1}{c|}{$\mathbf{A_g}$} & \multicolumn{1}{c|}{$A_g$} & \multicolumn{1}{c|}{$B_{1g}$} & \multicolumn{1}{c|}{$B_{2g}$} & $B_{3g}$ \\ \hline
\multicolumn{1}{c|}{$\mathbf{B_{1g}}$} & \multicolumn{1}{c|}{$B_{1g}$} & \multicolumn{1}{c|}{$A_g$} & \multicolumn{1}{c|}{$B_{3g}$} & $B_{2g}$ \\ \hline
\multicolumn{1}{c|}{$\mathbf{B_{2g}}$} & \multicolumn{1}{c|}{$B_{2g}$} & \multicolumn{1}{c|}{$B_{3g}$} & \multicolumn{1}{c|}{$A_g$} & $B_{1g}$ \\ \hline
\multicolumn{1}{c|}{$\mathbf{B_{3g}}$} & \multicolumn{1}{c|}{$B_{3g}$} & \multicolumn{1}{c|}{$B_{2g}$} & \multicolumn{1}{c|}{$B_{1g}$} & $A_g$ \\
\end{tabular}
\end{ruledtabular}
\end{table}

To characterize the symmetry of terms for which the independent-particle approximation is not directly applicable, the correlation tables describing the splitting of irreducible representations under a reduction of symmetry are presented in Table~\ref{tab:sym_3}. In particular, these correlation tables were used to present the data in Table~\ref{tab:table4}. Note that $D_{4h}$ contains two inequivalent classes of twofold axes perpendicular to the principal $C_4$ axis: $C_2'$, passing through the ligands, and $C_2''$, bisecting the ligand--metal--ligand angles. Retaining $C_2'$ upon lowering the symmetry to $D_{2h}$ corresponds to a rectangular distortion of the equatorial plane,
whereas retaining $C_2''$ corresponds to a rhombic (diagonal) distortion. These two embeddings are not equivalent and lead to different correlation tables, differing specifically in the images of $B_{1g}$ and $B_{2g}$ (the images of $A_{1g}$, $A_{2g}$, and $E_g$ are the same in both cases). Table~\ref{tab:sym_3} lists the correlation table for the rectangular convention, which is the one adopted throughout this work.

\begin{table}[h!]
\caption{\label{tab:sym_3} Correlation tables for the symmetry reductions $O_h \to D_{4h}$ and $D_{4h} \to D_{2h}$, the latter under the rectangular-distortion convention.}
  \begin{tabular}{cc}
    \hline\hline
    \multicolumn{2}{c}{$O_h \rightarrow D_{4h}$}       \\ \colrule
    $A_{1g}$ & $A_{1g}$          \\
    $A_{2g}$ & $B_{1g}$          \\
    $E_g$    & $A_{1g} \oplus B_{1g}$ \\
    $T_{1g}$ & $A_{2g} \oplus E_{g}$  \\
    $T_{2g}$ & $B_{2g} \oplus E_{g}$  \\ \colrule
    \multicolumn{2}{c}{$D_{4h} \rightarrow D_{2h}$}      \\ \colrule
    $A_{1g}$ & $A_g$             \\
    $A_{2g}$ & $B_{1g}$          \\
    $B_{1g}$ & $A_g$             \\
    $B_{2g}$ & $B_{1g}$          \\
    $E_g$    & $B_{2g} \oplus B_{3g}$ \\
    \hline\hline
  \end{tabular}
\end{table}

\bibliography{apssamp}

\end{document}